\documentclass[12pt,preprint]{aastex}
\usepackage{epsfig,natbib,lscape,fullpage,indentfirst,color}
\newcommand{\pdt}[1]{{{\partial #1}\over {\partial t}}}

\newcommand{\Inu}{I_{\nu}}
\newcommand{\kapa}{\kappa_{\star}\left(T,P\right)}
\newcommand{\kapp}{\kappa_P\left(T,P\right)}
\newcommand{\kapr}{\kappa_R\left(T,P\right)}

\newcommand{\er}{E_{R}}
\newcommand{\bbody}{B\left(T\right)}

\begin{document}
\title{Radiative Hydrodynamic Simulations of HD209458b: Temporal
Variability}

\author{Ian Dobbs-Dixon$^{1,2}$, Andrew Cumming $^1$, D.N.C
Lin $^{3,4}$}

\affil{$^1$Department of Physics, McGill University, Montreal, H3A 2T8
Canada}

\affil{$^2$Department of Astronomy, Box 351580, University of
  Washington, Seattle, WA 98195}

\affil{$^3$Department of Astronomy and Astrophysics, University of
California, Santa Cruz, CA 95064, USA}

\affil{$^4$Kavli Institute of Astronomy \& Astrophysics, Peking
University, Beijing, China}


\begin{abstract}
We present a new approach for simulating the atmospheric dynamics of
the close-in giant planet HD209458b that allows for the decoupling of
radiative and thermal energies, direct stellar heating of the
interior, and the solution of the full 3D Navier Stokes
equations. Simulations reveal two distinct temperature inversions
(increasing temperature with decreasing pressure) at the sub-stellar
point due to the combined effects of opacity and dynamical flow
structure and exhibit instabilities leading to changing velocities and
temperatures on the nightside for a range of viscosities. Imposed on
the quasi-static background, temperature variations of up to $15\%$
are seen near the terminators and the location of the coldest spot is
seen to vary by more than $20^{\circ}$, occasionally appearing west of
the anti-solar point. Our new approach introduces four major
improvements to our previous methods including simultaneously solving
both the thermal energy and radiative equations in both the optical
and infrared, incorporating updated opacities, including a more
accurate treatment of stellar energy deposition that incorporates the
opacity relevant for higher energy stellar photons, and the addition
of explicit turbulent viscosity.
\end{abstract}

\section{Introduction}
Close in gas-giant planets have by now become a familiar part of the
growing family of extrasolar planets. Their short period orbits and
proclivity for transiting has made them the target of numerous
observational campaigns and our knowledge of their structure and
composition has increased dramatically over the past few years. The
intense irradiation they receive from their host star dominates the
energy budget of the atmosphere and drives supersonic flows unlike any
seen in our own solar system. However, despite their prevalence and
important role in constraining a wide range of planetary models,
fundamental questions about the dynamical behavior of their
atmospheres remain, crucial for interpreting observations.

There have been a number of different groups working to understand the
dynamical redistribution of energy in close-in irradiated planets
utilizing a wide range of approaches. Because of the breadth of these
methods, interpreting the results and understanding their implications
can be quite confusing. Detailed comparisons of many of the models in
the literature have been presented previously \citep{dobbsdixon2008_1,
  showman2008_1, goodman2008} and we will not attempt to repeat those
discussions here but rather briefly place this present study in
context. In a general sense, the approaches taken to this problem can
be categorized using two criteria: the approach to radiation and the
approach to dynamics. Both components are extremely important to
include in any model. Energy transport, including radiation \emph{and}
advection, determines the distribution of thermal properties where as
pressure gradients and gravity regulate thermal currents throughout
the planet \citep{burkert2005}. A self consistent treatment of both
processes is required to determine the gas flow pattern and emerging
radiative spectra.

Approaches utilized for radiation to date include relaxation methods
({\it i.e.}  Newtonian heating) \citep{showman2002, showman2008_2,
  cooper2005, cooper2006, langton2007, langton2008, menou2008,
  rauscher2009}, kinematic constraints designed to represent incident
flux \citep{cho2003, cho2008, rauscher2008}, 3D flux-limited diffusion
\citep{burkert2005,dobbsdixon2008_1}, or 1D frequency dependent
radiative transfer \citep{showman2009}. Approaches to the dynamical
portion of the model have included solving the equivalent barotopic
equations \citep{cho2003, cho2008, rauscher2008}, the shallow water
equations \citep{langton2007, langton2008}, the primitive equations
\citep{showman2002, showman2008_2, showman2009, cooper2005,
  cooper2006, menou2008, rauscher2009}, and Euler's equations
\citep{burkert2005,dobbsdixon2008_1}.

Ideally we would integrate the full Navier-Stokes equations coupled to
a 3D frequency dependent radiative transfer model. However,
computational resources are limited, and simulations of this type take
considerable time. Thus, to our knowledge the approach of all groups
to date has been to choose one approach to dynamics and one to
radiation, necessarily compromising on one or the other. For the
results presented here we have chosen to utilize a two-frequency, 3D
flux-limited model coupled to the full 3D compressible Navier-Stokes
equations. Our goal in this study is to understand the details of
atmospheric dynamics and energy transfer in highly irradiated
environments which are not spherically symmetric. Our new approach to
the radiation flux reproduces many of the general features of
frequency-dependent approaches. Most importantly, it allows for energy
deposition at a depth corresponding to the optical photosphere. The
depth of deposition controls much of the subsequent dynamics and a
bolometric approach to the energy transfer is all that is needed.

The plan of the paper is as follows; Section (\ref{sec:methods})
describes our numerical approach, concentrating on the improvements to
our methods from our previous paper
\citep{dobbsdixon2008_1}. Primarily, this improvement lies in the
treatment of the energy equation in which we track the energetic
components separately. In addition, we use updated opacities
calculated for infrared and optical components independently. In
Section (\ref{sec:results}) we present results for the well known
HD209458b. We begin by illustrating the radiative solution,
reproducing many of the important features of the pressure-temperature
profiles from more detailed 1D models. We then present a number of
aspects of the underlying flow and atmospheric structure for a range
of viscosities. Viscosity plays an important role in both regulating
flow velocities and providing viscous heating. For a range of
viscosities, our simulations exhibit dynamical variations that give
rise to changing temperature distributions across the
photosphere. Section (\ref{sec:timedep}) presents several aspects of
this exo-weather, in particular changes we expect to see from multiple
observations. Finally, several questions regarding the nature of the
energy flow throughout the atmosphere have recently been posed by
\citet{goodman2008}. To address this, in Section (\ref{sec:fluxes}) we
discuss the role of viscosity in determining the entropy distribution
throughout the atmosphere, and the relative fluxes of kinetic energy,
enthalpy, and radiation from day to nightside. We conclude with a
discussion in Section (\ref{sec:conc}).

\section{Model Description}
\label{sec:methods}
We model the planetary atmosphere using a three-dimensional radiative
hydrodynamical model in spherical coordinates
$\left(r,\phi,\theta\right)$. In addition to the numerical techniques
utilized in solving the hydrodynamical portion of the code described
in \citet{dobbsdixon2008_1}, we have added explicit viscosity terms
to our equations in the form described by
\citet{kley1987}. Significant changes to the radiative component of
our atmospheric model have also been implemented. In addition to
updated opacities \citep{sharp2007}, we have decoupled the radiative
and thermal energy components and introduced a multi flux channel
approach for the radiative energy component in a manner similar to the
approach of \citet{howell2003}.

The fully compressible three dimensional Navier-Stokes equation for
the motion of the fluid is given by
\begin{equation}
\pdt{{\bf u}} + \left({\bf u}\cdot\nabla\right) {\bf u} = -
\frac{\nabla{P}}{\rho} + {\bf g} -2{\bf \Omega\times u} - \nonumber
{\bf\Omega\times}\left({\bf\Omega\times r}\right) +\nu\nabla^2{\bf u}
+\frac{\nu}{3}\nabla\left(\nabla\cdot{\bf u}\right)
\label{eq:momentum}
\end{equation}
where {\bf u} is the three-dimensional velocity, $\rho$ and $P$ are
the gas density and pressure, and $\Omega$ is the rotation
frequency. The gravitational acceleration, ${\bf g}$ is taken here to
be purely radial. $\nu=\eta/\rho$ is the constant kinematic viscosity,
proportional to $\eta$, the dynamic viscosity. We have neglected the
coefficient for the bulk viscosity. For further details on the
treatment of the viscous terms we direct the reader to
\citet{kley1987}. Explicit viscosity is supplemented by an additional
artificial viscosity for the accurate treatment of shocks. The
functional form of this term is identical to that in the ZEUS code
\citep{stone1992}. Equation (\ref{eq:momentum}) is supplemented by the
continuity equation, given by
\begin{equation}
\pdt{\rho} + \nabla\cdot\left(\rho{\bf u}\right) = 0.
\label{eq:continuity}
\end{equation}

The equation for internal energy of the gas can be written as
\begin{equation}
\left[ \pdt{\epsilon} + ({\bf u}\cdot\nabla) \epsilon \right] = - P \,
\nabla \cdot {\bf u} - \sum_i \rho c
\kappa_{P,i}\left(\frac{j_i}{c\kappa_{P,i}} - E_{R,i} \right) + D_\nu,
\label{eq:energy_first}
\end{equation}
The quantity proportional to $j_i$ represents the photons added to the
i-th radiation energy density, while the term on the right-hand side
proportional to $E_{R,i}$ represents those absorbed by the
gas. $\kappa_{P,i}$ is the Planck opacity, defined below. Ignoring
non-coherent scattering, these terms represent either impinging
radiation or interactions with the surrounding gas. In the above
$\epsilon=c_{v} \rho T$ is the internal energy density, $c_v$ is the
specific heat, and $D_\nu=\left({\bf S}\nabla\right){\bf v}$ is the
viscous dissipation function. The components of the viscous stress
tensor ${\bf S}$ are detailed in \citet{kley1987} or
\citet{milhalas1984}. With the exception of the isotropic assumption
for $j_{\nu}$ and $\kappa_{\nu}$ (used during the angular integration
of Equation (\ref{eq:radiation}) below) and neglecting scattering,
this set of equations is completely general and can be expanded to an
arbitrary number of frequency bins.

In addition to the internal energy of the gas we must also derive an
expression for the radiation energy. For this consider a simplified
frequency dependent model with two frequency ranges. We can write the
radiative transfer equation as
\begin{equation}
\frac{1}{c}\pdt{\Inu} + \hat{\bf k}\cdot\nabla\Inu =
\rho\left(\frac{j_{\nu}}{4\pi} +
\kappa_{\nu}^{scat}\Phi_{\nu}^{scat}\right) - \rho\kappa_{\nu}\Inu.
\label{eq:radiation}
\end{equation}
In the above $\Inu$ is the specific intensity, $j_{\nu}$ is the
emissivity, and $\kappa_{\nu} = \kappa_{\nu}^{abs} +
\kappa_{\nu}^{scat}$ is the total opacity; the sum of the absorption
and scattering opacities. The intensity of photons scattered into the
beam is given by $\Phi_{\nu}^{scat}=\int\phi_{\nu}\left({\bf
k^\prime},{\bf \hat{k}}\right)\Inu\left({\bf
k^\prime}\right)d\Omega^\prime$, where $\phi_{\nu}\left({\bf
k^\prime},{\bf \hat{k}}\right)$ is the scattering probability density
between the ${\bf \hat{k}}$ and ${\bf k^\prime}$ beams. Integrating
Equation (\ref{eq:radiation}) over angle, we define the moments as
$\left(cE_{R,\nu},{\bf F}_{\nu}\right) = \int\left(1,\hat{\bf
k}\right)\Inu d\Omega$, where $E_{R,\nu}$ and ${\bf F_{\nu}}$ are the
frequency dependent radiation energy density and flux respectively.

We now identify the first frequency interval with the incoming stellar
radiation. This is characterized by the larger $T_{\star}$, and peaks
in the optical. The second frequency interval then represents the
local re-radiated energy, characterized by the local temperature
$T$. This peaks in the infrared for the problem of interest
here. Addressing the impinging stellar photons first (group $1$) we
assume that the atmospheric gas does not emit in this wave-band,
implying $j_1=0$. To find an expression for $E_1$, the stellar energy
absorbed by the gas, we return to the full radiative transfer
equation. Allowing only absorption, and neglecting temporal
variations, Equation (\ref{eq:radiation}) for the intensity of the
stellar radiation can be written simply as
\begin{equation}
\frac{dI_\star}{dr} = -\rho\kappa_\star I_\star,
\end{equation}
with solution $I_\star = I_{\star,0}\exp\left[-\tau_\star\right]$. We
have denoted this frequency interval with the subscript $\star$. The
stellar heating term in Equation (\ref{eq:energy_first}) can then be
written as
\begin{equation}
E_{R,1} = E_\star = \frac{1}{c}\int I_\star d\Omega =
\frac{1}{c}I_{\star,0}e^{-\tau_\star} = W a
T_{\star}^4e^{-\tau_\star},
\label{eq:E1}
\end{equation}
where we have explicitly assumed that this radiation is {\it not}
isotropic but rather flows radially from the central star. The factor
$W$ accounts for the attenuation of the stellar energy and is
inversely proportional to the square of the semi-major axis. The
optical depth for incoming stellar radiation is given by
$\tau_{\star}$.

We now return to Equation (\ref{eq:radiation}) for the local,
re-processed radiation (frequency interval $2$), which peaks in the
infrared for this problem. Integrating Equation (\ref{eq:radiation})
over angle and frequency, dropping the subscript $2$, assuming the gas
radiates this energy isotropically and is in local thermodynamic
equilibrium ({\it i.e.} $j=B\kappa$), the radiative energy component
can be expressed as,
\begin{equation}
\pdt{\er} + \nabla \cdot {\bf F} = \rho \kapp\left[\bbody-c\er\right].
\label{eq:radenergy}
\end{equation}
In the above integration we have assumed that $j_{\nu}$ and
$\kappa_{\nu}$ are isotropic. The local radiative energy density $\er$
is evolved independently from the thermal component using this
equation, with the term $\bbody=4\sigma T^4$ linking it to the gas.

Combining Equations (\ref{eq:energy_first}) and (\ref{eq:E1}), the
final thermal energy equation is given by
\begin{equation}
\left[ \pdt{\epsilon} + ({\bf u}\cdot\nabla) \epsilon \right] = - P
\nabla \cdot {\bf u} - \nonumber \rho \kapp\left[\bbody - c\er \right] +
\rho\kapa F_{\star} e^{-\tau_\star} + D_v.
\label{eq:thermalenergy}
\end{equation}
The second term on the left accounts for the advection of thermal
energy throughout the planet, the first term on the right is the
compressional heating term, the term proportional to $\bbody-c\er,$
represents the exchange of energy between the thermal and radiative
components through the emission and absorption of the low energy
photons, and $\rho\kapa F_{\star} e^{-\tau_{\star}}$ represents the
higher energy stellar photons absorbed by the gas. The stellar flux at
the planet's dayside surface will be
\begin{equation}
F_{\star} = \sigma T_{\star}^4\left(\frac{R_{\star}}{a}\right)^2
\left[\cos\left(\theta\right)\cos\left(\phi\right)\right],
\label{eq:Fstar}
\end{equation}
where $\theta$ and $\phi$ are latitude and longitude on the planet's
surface.

There are two final ingredients necessary for solving the energy
balance in the atmosphere; a prescription for the opacity and a
closure relation linking the flux ${\bf F}$ back to the radiation
energy density. Here we utilize the three dimensional flux-limited
diffusion (FLD) approximation of \citet{levermore1981}, where
\begin {equation}
{\bf F} = - \lambda {\frac{c}{\rho\kapr}} \nabla \er.
\label{eq:flux}
\end {equation}
The Rosseland opacity is given by $\kapr$, and $\lambda$ is a
temporally and spatially variable flux limiter providing the closure
relationship between flux and radiation energy density. The functional
form of $\lambda$ is given by
\begin{equation}
\lambda = \frac{2+R}{6+3R+R^2},
\label{eq:lambda}
\end {equation}
where
\begin {equation}
R = \frac{1}{\rho \kapr} \frac{| \nabla \er |}{\er}.
\label{eq:Rdef}
\end {equation}
FLD has been utilized in a wide range of astrophysical applications
and allows for the simultaneous study of optically thick and optically
thin gas, correctly reproducing the limiting behavior of the
radiation at both extremes. In the optically thick limit the radiation
energy can be expressed as $E=aT^4$, and the flux becomes the standard
radiative diffusion equation,
\begin {equation}
{\bf F} = - \frac{4acT^3}{3\rho\kapr} \nabla T.
\label{eq:Fthick}
\end {equation}
In the optically thin streaming limit, Equation (\ref{eq:flux})
becomes
\begin {equation}
|{\bf F}| = c\er.
\label{eq:Fthin}
\end {equation}
Between these limits, Equation (\ref{eq:lambda}) approximates the full
wavelength dependent radiative transfer models of
\citet{levermore1981}.

There are three frequency averaged opacities appearing in Equations
(\ref{eq:radenergy}), (\ref{eq:thermalenergy}), and (\ref{eq:flux}):
$\kapp$, $\kapa$, and $\kapr$. The importance of differing absorption
opacity and emissivity was first noted by \citet{hubeny2003} but not
widely appreciated until the discovery of the stratosphere of
HD209458b \citep{knutson2008}. The modeling of HD209458b \citep[{\it
    e.g.}]{burrows2007} suggests a thermal inversion in the upper
atmosphere largely due to the differing opacities. Here we include
these effects within the framework of frequency averaged opacities. 3D
frequency dependent radiative transfer coupled to the full 3D
Navier-Stokes equations would be the ideal tool for these
calculations; unfortunately, to our knowledge no such calculations
exist to date. Although our models do not contain the full
wavelength dependent opacities of one-dimensional radiative models,
with the addition of multiple average opacities and the radiative
energy equation we are able to capture many of the essential features
that appear in such models. The energetics powering the underlying
dynamics is well represented by an average approach. Moreover, this
approach allows us to still solve the full three dimensional Navier
Stokes Equations with reasonable computational requirements. Our
previous atmospheric models \citep{dobbsdixon2008_1} not only took the
radiative and thermal temperatures to be equal but also assumed that
all three of these opacities were equivalent. Of particular
importance, our new approach modifies the radial location that the
stellar energy is initially deposited and the winds are launched. The
unfortunate side-effect is that we cannot produce detailed spectra
directly from these models. However, it is possible to post-process
the pressure-temperature profiles derived here using a one-dimensional
radiative transfer code, which will be presented elsewhere.

Planck and Rosseland mean opacities are defined in the usual manner,
but the temperature at which the spectra is evaluated is critical. The
local Planck mean for the low-energy photon group is given by
\begin{equation}
\kapp = \frac{\int \kappa_{\nu,ns}\left(T,P\right)
B_\nu\left(T\right)d\nu}{\int B_\nu\left(T\right)d\nu},
\label{eq:kappaP}
\end{equation}
while the Planck mean for the high-energy photon group is defined as
\begin{equation}
\kapa = \frac{\int \kappa_{\nu,ns}\left(T,P\right)
J_\nu\left(T_{\star}\right)d\nu}{\int J_\nu\left(T_{\star}\right)d\nu}.
\label{eq:kappaA}
\end{equation}
The frequency dependent opacity is given by $\kappa_{\nu}$, and the
subscript {\it ns} indicates scattering processes are neglected when
calculating the wavelength dependent opacities. In principle, the
impinging radiation can differ from a black-body, but for our purposes
we set
$J_\nu\left(T_{\star}\right)=B_\nu\left(T_{\star}\right)$. Finally,
the Rosseland mean for the low-energy group is given by
\begin{equation}
\kapr^{-1} = \frac{\int \kappa^{-1}_{\nu,s}\left(T,P\right)
  \frac{\partial B_\nu\left(T\right)}{dT}d\nu}{\int \frac{\partial
  B_\nu\left(T\right)}{dT}d\nu}.
\label{eq:kappaR}
\end{equation}
Here wavelength dependent opacities include the effect of scattering
(subscript {\it s}). In the calculations presented here we use the
wavelength dependent opacities as described in \citet{sharp2007}

\citet{burrows2008} find that matching the spectral data from
HD209458b requires the addition of a supplemental, unknown uniform
opacity source of $\kappa_{e}=0.1 \mathrm{cm^2/g}$ at pressures below
$0.03$ bars. The source of this additional opacity is currently
unknown. Some of the leading contenders are titanium oxide (TiO) and
vanadium oxide (VO) \citep{hubeny2003}. However, though TiO and VO may
indeed be the dominate source of opacity in the upper atmosphere at
high temperatures ($T \gtrsim 1500$K), its abundance remains an
outstanding question. \citet{spiegel2009} find, in the absence of
abnormally large vertical mixing coefficients, it is likely that VO
settles out of the upper atmosphere and does not play a role in
producing thermal inversions. \citet{zahnle2009} suggest $S_2$ and
$S_3$ as another potential source of high altitude absorption. Rather
then introduce the complex temperature and pressure dependence of an
unknown amount of TiO, VO, $S_2$, or $S_3$, our opacities \emph{do
not} contain these explicit compounds. Instead we follow
\citet{burrows2008} and augment the Planck opacities in the upper
atmosphere with the straightforward $\kappa_{e}$ parameterization. The
value of $\kappa_{e}$ is simply a fit to current observations and is
easily modified for studying other situations.

\section{Energy Flow Throughout the Planetary Atmosphere}
\label{sec:results}
In this section we present simulations of the full
radiative-hydrodynamical solutions of Equations (\ref{eq:momentum}),
(\ref{eq:continuity}), (\ref{eq:radenergy}), and
(\ref{eq:thermalenergy}) for conditions meant to represent the
close-in giant planet HD209458b. To this end, the orbital and
(assumed) rotation periods of the planet are $P=3.52\mathrm{ days}$,
corresponding to a semi-major axis of $0.047$ AU. The host star is
taken to have a radius of $1.15 R_{\odot}$ and a temperature of
$6030$K \citep{mazeh2000}. Although observable parameters are chosen
from the literature, we have not attempted to tune the models to fit
observed spectra or light-curves, with the exception of the extra
opacity source added (see Section (\ref{sec:methods})). Our goal here
is to explore general properties of atmospheric flow; specific models
with tuned parameters will be presented elsewhere. These results can
be considered representative of flows as you would see on planets
similar to HD209458b.

We solve the problem in 3D spherical coordinates with a standard
resolution of $(N_r,N_{\phi},N_{\theta}) = (60,160,64)$. Due to the
decreasing grid size near the pole, and the associated Courant
time-step constraint, we limit the latitudinal extent of our grid to
$\pm 70^o$. These boundaries are impenetrable and slip-free, and the
flux is set to zero. The role of these artificial poles on the overall
flow is difficult to characterize and is an unfortunate side effect of
the chosen coordinate system. We have run tests changing
$\left|\theta_{max}\right|$ by $10^{\circ}$ and have observed little
effect on the overall atmospheric structure. However, this test is not
conclusive. A more concrete discussion of the effect of this boundary
requires eliminating this boundary by means of a 'polar-patch' which
we hope to include in future models.

All simulations are initialized with the static one-dimensional
equilibrium pressure-temperature profiles as calculated in
\citet{burrows2007}. As discussed below in Section (\ref{sec:rad}), we
then solve the radiative portion of the energy equations, subject to
the non-symmetric stellar irradiation until the simulation reaches
radiative equilibrium. Dynamical results, discussed in Section
(\ref{sec:mean}) are initialized with this static
radiative-equilibrium model. The gas at the inner boundary is assumed
to be optically thick ($\er=aT^4$) and its properties spherically
symmetric while the boundary is assumed to be impenetrable and slip
free. We set a spherically symmetric radiative flux and calculate the
temperature gradient with Equation (\ref{eq:Fthick}). The pressure at
the inner boundary is set well within the spherically symmetric
convective interior, allowing us to both avoid any influence of the
choice of inner velocity boundary conditions on the upper atmospheric
flow and allow for the interaction between the convective interior and
irradiation induced dynamics. A set of less restrictive inner boundary
conditions are needed to determine the influence of stellar
irradiation on convection, cooling, and the quasi hydrostatic
contraction rates of hot Jupiters.  Such simulations will be presented
elsewhere. A movable outer boundary ({\em not} associated with either
the optical or infrared photospheres) is located at a density of
$10^{-9} \mathrm{g/cm^3}$, through which we assume a constant outward
radiative flux ($dF_R/dr = 0$) through an optically thin gas, given by
Equation (\ref{eq:Fthin}). Given the negligible effect of dynamics at
the lowest pressures, the gas temperature at the outer boundary is set
to its thermal steady-state value with Equation (\ref{eq:Tanalytic}),
given below.

\subsection{Radiative Solution}
\label{sec:rad}
\begin{figure}
\plotone{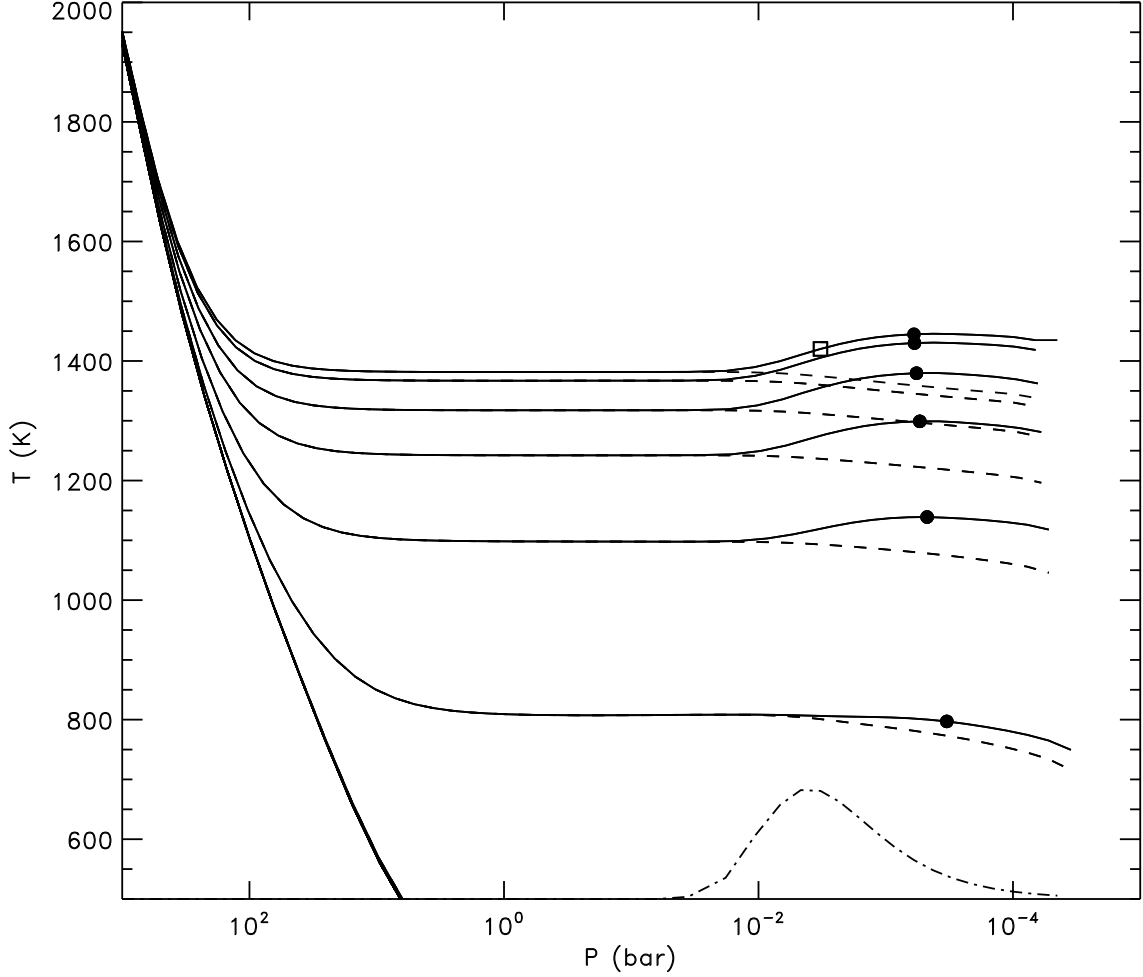}
\caption{The temperature as a function of pressure
for a purely radiative solution (solid line) from the sub-stellar
point to anti-stellar point. The dashed line shows $(\er/a)^{1/4}$ and
illustrates the decoupling of thermal and radiative energies near the
photosphere. Solid points denote the location of the infrared
photosphere, while the square shows the location of the optical
photosphere on the dayside. Also shown (dot-dash line) is the extent
of the stellar heating $\rho\kapa F_{\star} e^{-\tau_\star}$.}
\label{fig:radPT}
\end{figure}

Before including the dynamical portion of the code, we first allow the
simulation to come to radiative equilibrium. This is accomplished by
solving the energy equations given by Equations (\ref{eq:radenergy})
and (\ref{eq:thermalenergy}) with ${\bf u}=0$. The temperature
profiles are shown in Figure (\ref{fig:radPT}). The solid line denotes
the gas temperature $T$ and the dashed line denotes the radiative
energy $\er$. Decoupling the energy into thermal and radiative
components is a crucial element to account for when studying an
irradiation induced upper atmospheric temperature inversion, and is
clearly evident near the stellar photosphere where $\er$ remains
roughly constant. Also shown in Figure (\ref{fig:radPT}) are the
locations of the infrared and optical photospheres and the (scaled)
direct stellar heating term $\rho\kapa F_{\star}e^{-\tau_\star}$. The
extent of the temperature inversion is clearly associated with the
deeper optical photosphere and the resulting extent of the heating
term. Our radiative pressure-temperature profiles are quite similar to
those in \citet{burrows2007}, with the exception of the ad-hoc energy
sink they add to the dayside at high pressures to mimic the effects of
dynamics. This energy sink decreases the temperature at depth,
producing a secondary inversion in their models. This feature is seen
self-consistently in our dynamical simulations (Section
(\ref{sec:mean})).

To understand the behavior of the temperature at low optical depths
consider the steady-state behavior of Equation
(\ref{eq:thermalenergy}) for zero velocity. In this case the gas
temperature can be expressed as
\begin{equation}
T = \left[\frac{\er}{a} + \gamma^4\frac{F_{\star}}{4\sigma}
  e^{-\tau_{\star}}\right]^{1/4},
\label{eq:Tanalytic}
\end{equation}
Local equilibrium between absorption and emission requires that the
gas temperature be greater then $\left(\frac{\er}{a}\right)^{1/4}$. In
the above, $\gamma$ is the ratio of Planck opacities evaluated for the
stellar and local fluxes as defined by \citet{hubeny2003} as
\begin{equation}
\gamma^4 \equiv \frac{\kapa}{\kapp}.
\label{eq:gamma}
\end{equation}
The role that the two opacities play in creating a temperature
inversion is apparent in Equation (\ref{eq:Tanalytic}), the nature of
which depends sensitively on the behavior of $\gamma$ with depth; an
inversion requires $\gamma^4 e^{-\tau_{\star}}$ to increase outward,
while a temperature peak in the outer atmosphere (seen in some
simulations not presented here) requires a peak in this quantity. If
the opacity is such that regions exist above the optical photosphere
with $\gamma>1$, the gas temperature can easily become super-heated,
exceeding the equilibrium temperature calculated from Equation
(\ref{eq:Fstar}).

\subsection{Mean Dynamical Solution}
\label{sec:mean}
In this section we present the full radiative hydrodynamical
simulations for HD209458b. All dynamical simulations are initialized
with the zero velocity radiative equilibrium results discussed in
Section (\ref{sec:rad}). Fluid velocities are initially artificially
damped then slowly ramped to their full values to avoid spurious
oscillations or shocks associated with the initialization. As
discussed below, several aspects of the flow and temperature
distribution have significant time varying components. Results shown
in this section are averaged over approximately one day, in hopes of
representing mean flow properties. Results presented here are taken
from simulations that have run for a minimum of $100$ simulated
days. Mean temperature and velocity distributions appear to have
stabilized by this point. Section (\ref{sec:timedep}) presents the
superimposed time-varying characteristics.

To explore the role of viscosity we have repeated our analysis for a
range of kinematic viscosities, shown in Table (\ref{table:one}). We
will refer to runs with different viscosities using the notation
S1-S4. In the absence of more detailed knowledge of the detailed
dissipative processes ({\it e.g.} small scale instabilities or
turbulence) we take the kinematic viscosity to be a constant. Also
listed in Table (\ref{table:one}) is the characteristic value of
$\alpha_{eff,ph}=\nu/\left(c_sH\right)$ at the photosphere for each
simulation, where $H_p$ is the pressure scale-height at the
photosphere. Although we do not utilize the $\alpha$-prescription, it
is a useful way to normalize our chosen viscosity values. Our
viscosity is assumed to be ultimately related to the generation of
turbulence and the cascade of this energy to small length
scales. Turbulence may be generated through many processes including
various shear instabilities (associated for example with overturning
Kelvin-Helmholtz instabilities), waves acting throughout the atmosphere
(both inertial waves in the convective regions and Hough modes in the
radiative regions \citep{ogilvie2004}), and shocks generated within
the interacting flow. Given the limited resolution achievable in
dynamical simulations, the cascade of turbulence to scales smaller
than our grid size and the subsequent damping of that energy must
necessarily be represented by a sub-grid model. \citet{penev2008} have
recently shown that turbulent flows can be well represented by such an
effective viscosity coefficient, such as in the final two terms in
Equation (\ref{eq:momentum}). Given the uncertainty of the sub-grid
physics, we have chosen to run models for a range of $\nu$. As is
explained below the values bracket a critical viscosity $\nu_{crit}$,
for which we expect viscous effect to be (sub)dominate. In addition,
there is some amount of observational motivation for this choice from
observations of Jupiter's photochemistry. \citet{moses2005} show that
the eddy diffusion coefficient for particles at pressures of $10^{-4}
\mathrm{bars}$ to be $10^5-10^6 \mathrm{cm^2/s}$. Presumably momentum
diffusion will be more effective than species diffusion.

\begin{table}
\begin{center}
\begin{tabular}{|c|c|c|c|c|}
\hline
Simulation & $\nu$ ($\mathrm{cm^2/s}$) & $\alpha_{eff,ph}$ & $H_{p,ph}$ (km) & peak $v_{\phi}$ (km/s)\\
\hline
\hline
$S1$ & $10^{12}$ & $10^{-1}$ & 330 & 0.8 \\
\hline
$S2$ & $10^{10}$ & $10^{-3}$ & 360 & 4.5 \\
\hline
$S3$ & $10^{8}$  & $10^{-5}$ & 360 & 5.3 \\
\hline
$S4$ & $10^{4}$  & $10^{-9}$ & 360 & 5.7 \\
\hline
\end{tabular}
\end{center}
\caption{Values of kinematic viscosity used for the simulations
  presented here. For reference we also quote an average effective
  alpha-parameter and pressure scale-height all calculated at the
  photosphere. Peak velocities, shown in the last column, increase
  with decreasing viscosity.}
\label{table:one}
\end{table}

\begin{figure}
\includegraphics[height=7.0cm,width=8.0cm]{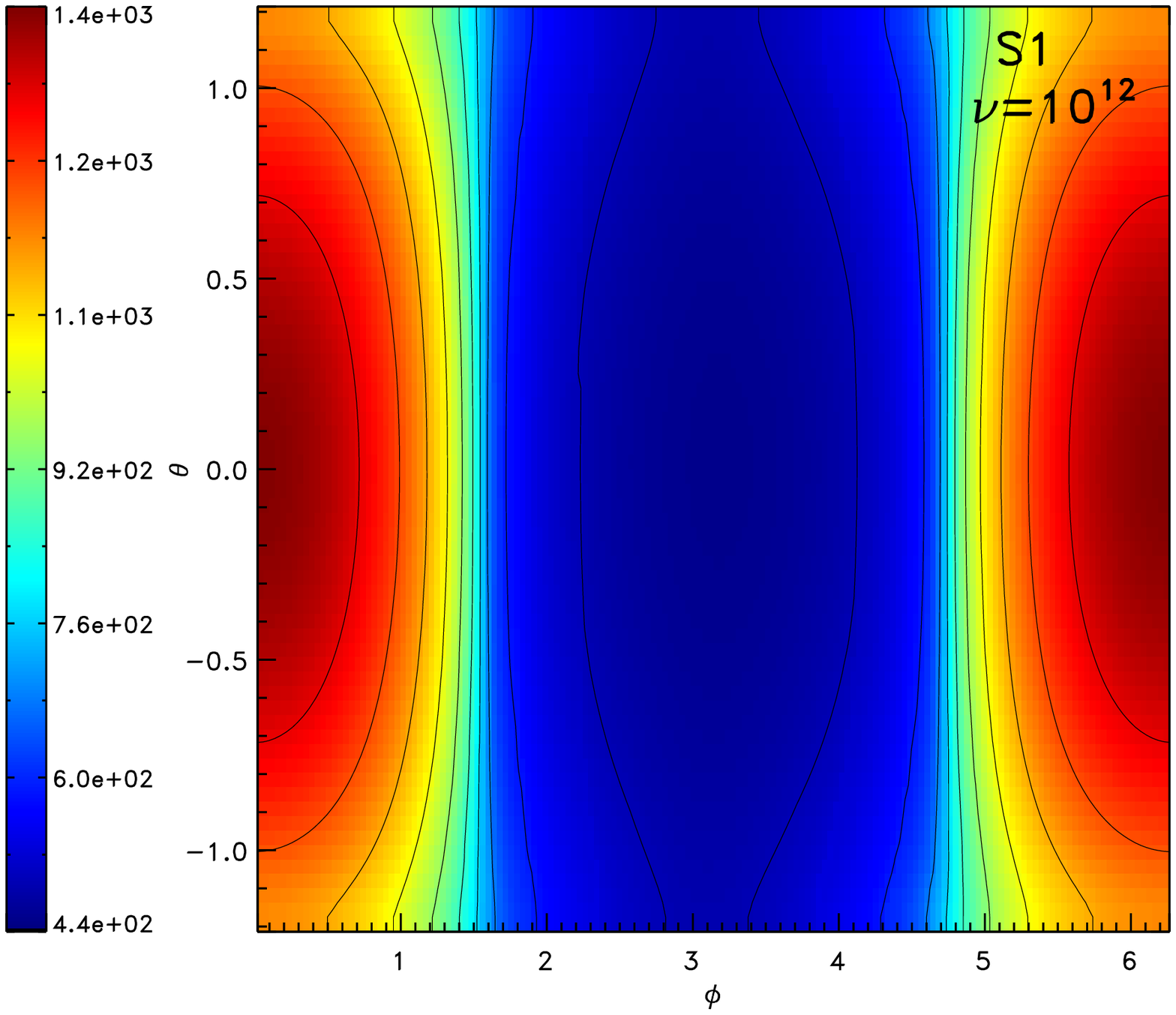}
\hfill
\includegraphics[height=7.0cm,width=8.0cm]{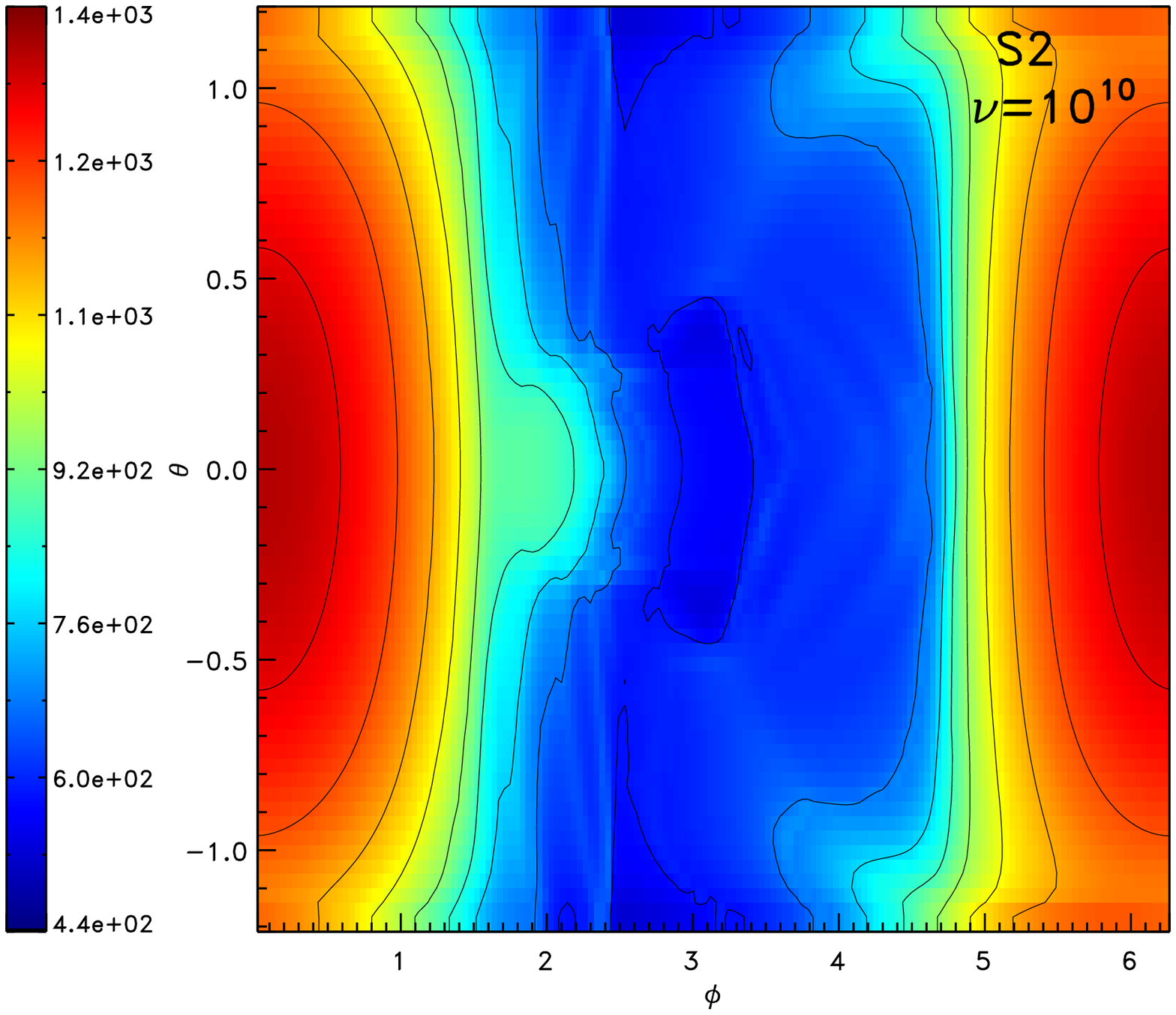}
\includegraphics[height=7.0cm,width=8.0cm]{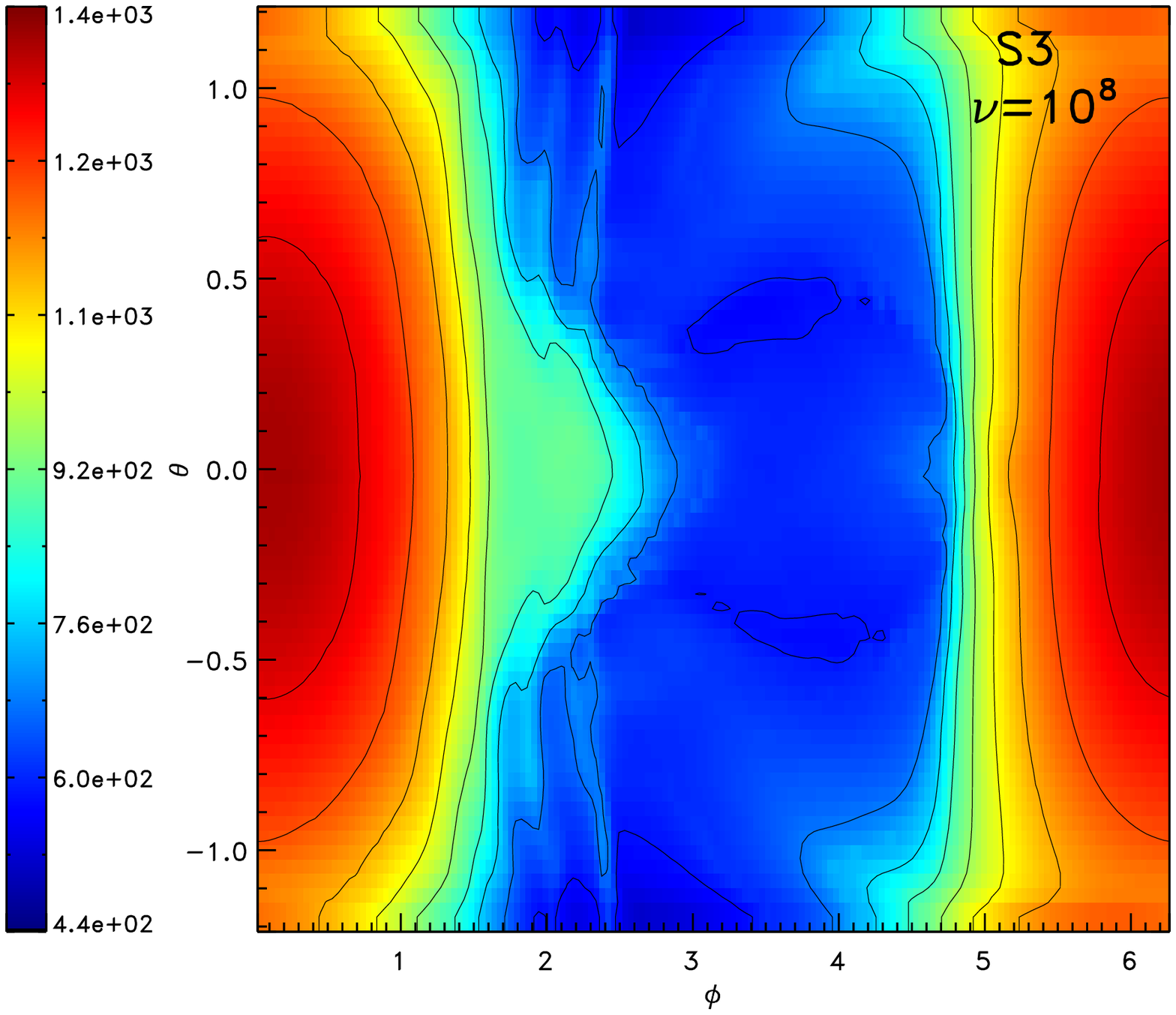}
\hfill
\includegraphics[height=7.0cm,width=8.0cm]{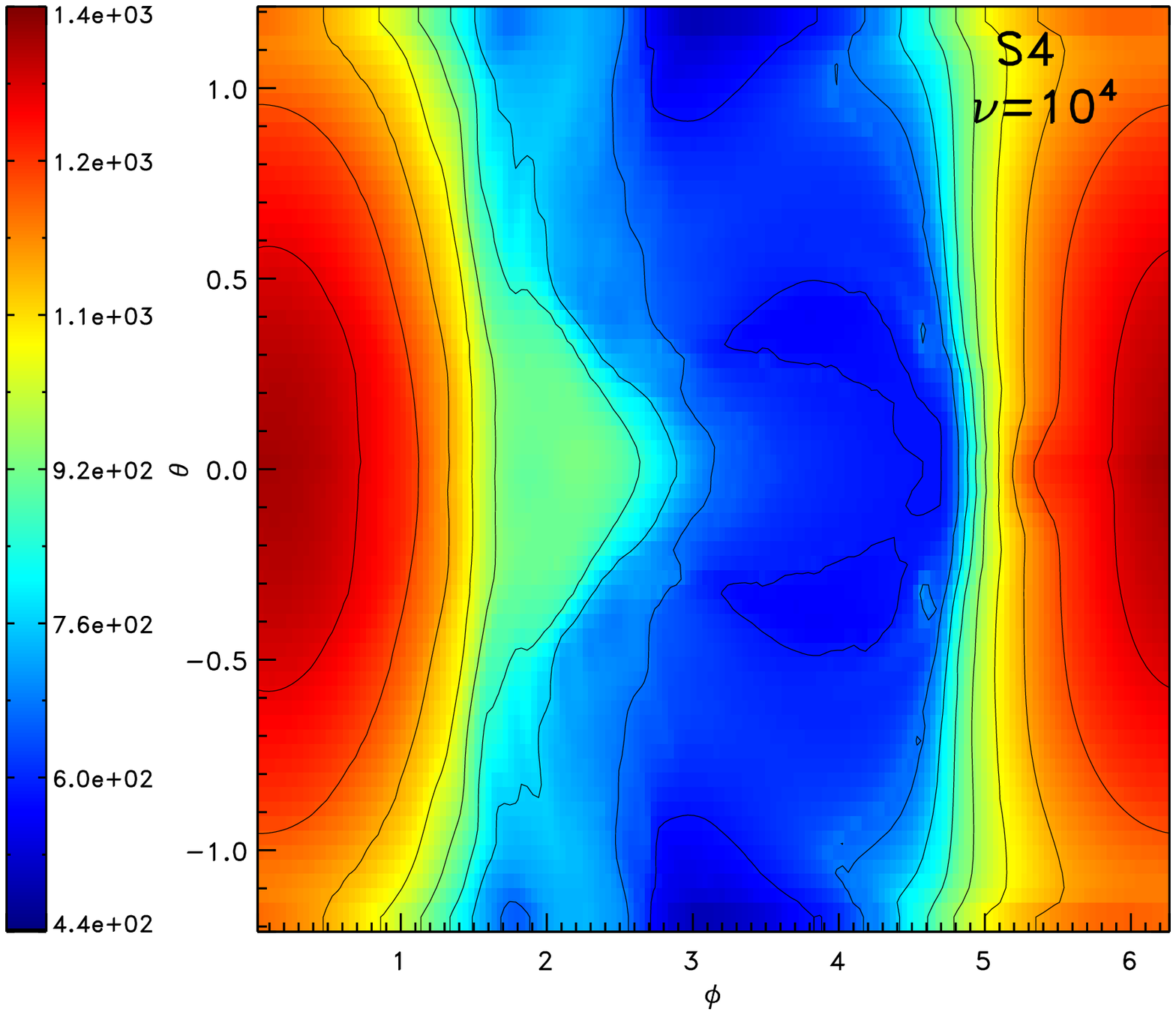}
\caption{The temperature at the photosphere for simulations with
  varying viscosities. Viscosities for upper-left, upper-right,
  lower-left, and lower-right are $10^{12}$, $10^{10}$, $10^{8}$, and
  $10^{4} \mathrm{cm^2/s}$ respectively. The center of each graph is
  located at the anti-stellar point
  $\left(\phi,\theta\right)=\left(\pi,0\right)$ and the equator
  ($\theta=0$) runs horizontally through the center.  For $10^{12}
  \mathrm{cm^2/s}$, the lack of advection keeps the entire nightside
  cool, while for the other simulations advection significantly alters
  the temperature structure near the terminators ($\pi/2$ and
  $3\pi/2$) and across the nightside.}
\label{fig:Tph}
\end{figure}

Figure (\ref{fig:Tph}) shows the temperature at the infrared
photosphere for simulations with a range of viscosities. Temperatures
range from $1400$K to $440$K at the photosphere. The hottest
photospheric point lies slightly east ($\phi > 0$) of the sub-stellar
point in all simulations, while the coolest points are at higher
latitudes (both above and below the equator) on the nightside. Along
the equator, the location of the coolest point depends on viscosity,
lying east of the anti-stellar point ($\phi > \pi$) for the low
viscosity runs S3 and S4. For S2, viscous heating associated with the
converging flows drives oscillations in the location of the coldest
point (see Section (\ref{sec:timedep})). The highest viscosity run
(S1) exhibits very little flow on the nightside and the coldest point
remains very near the anti-stellar point.

\begin{figure}
\includegraphics[height=7.0cm,width=8.0cm]{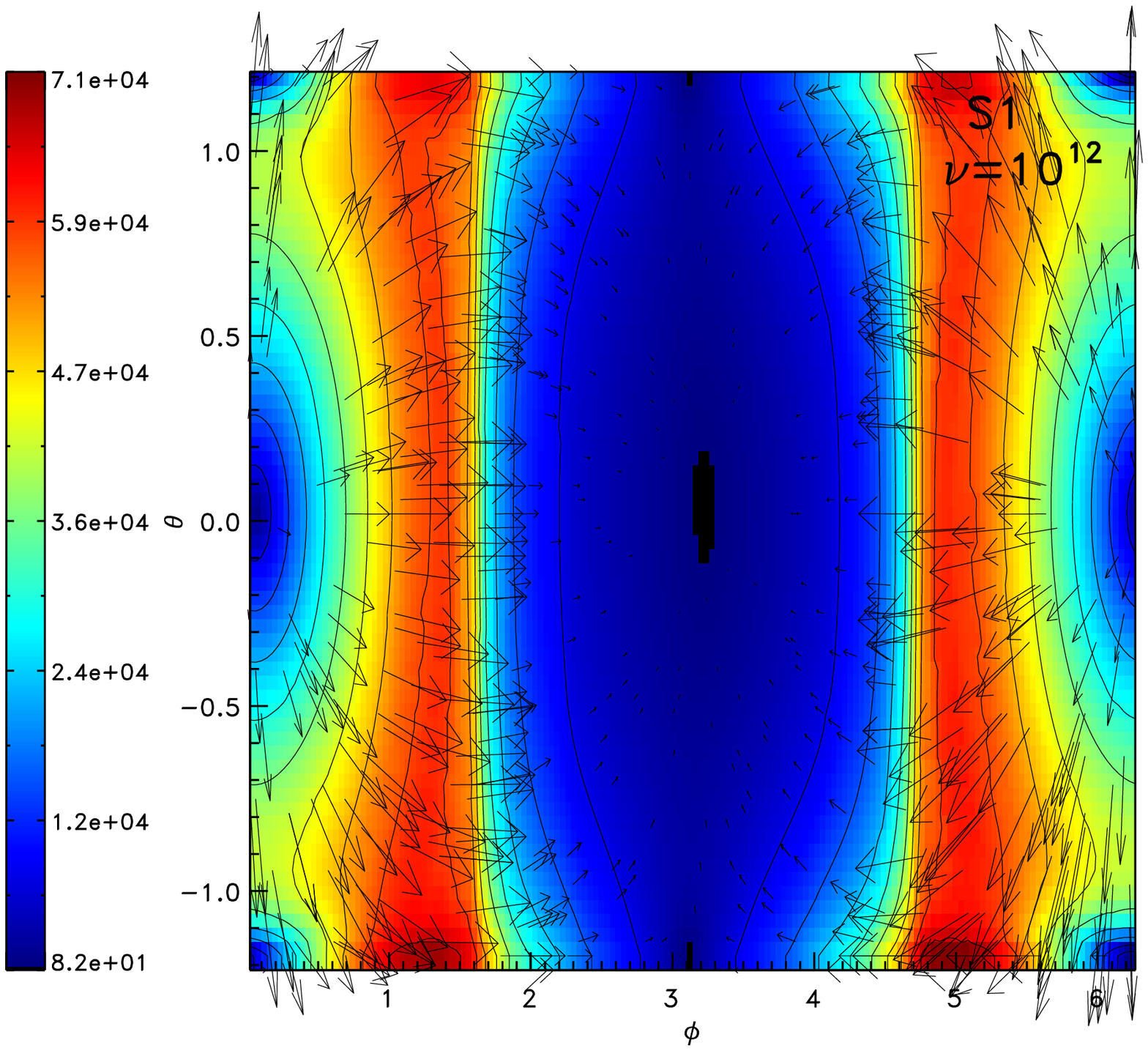}
\hfill
\includegraphics[height=7.0cm,width=8.0cm]{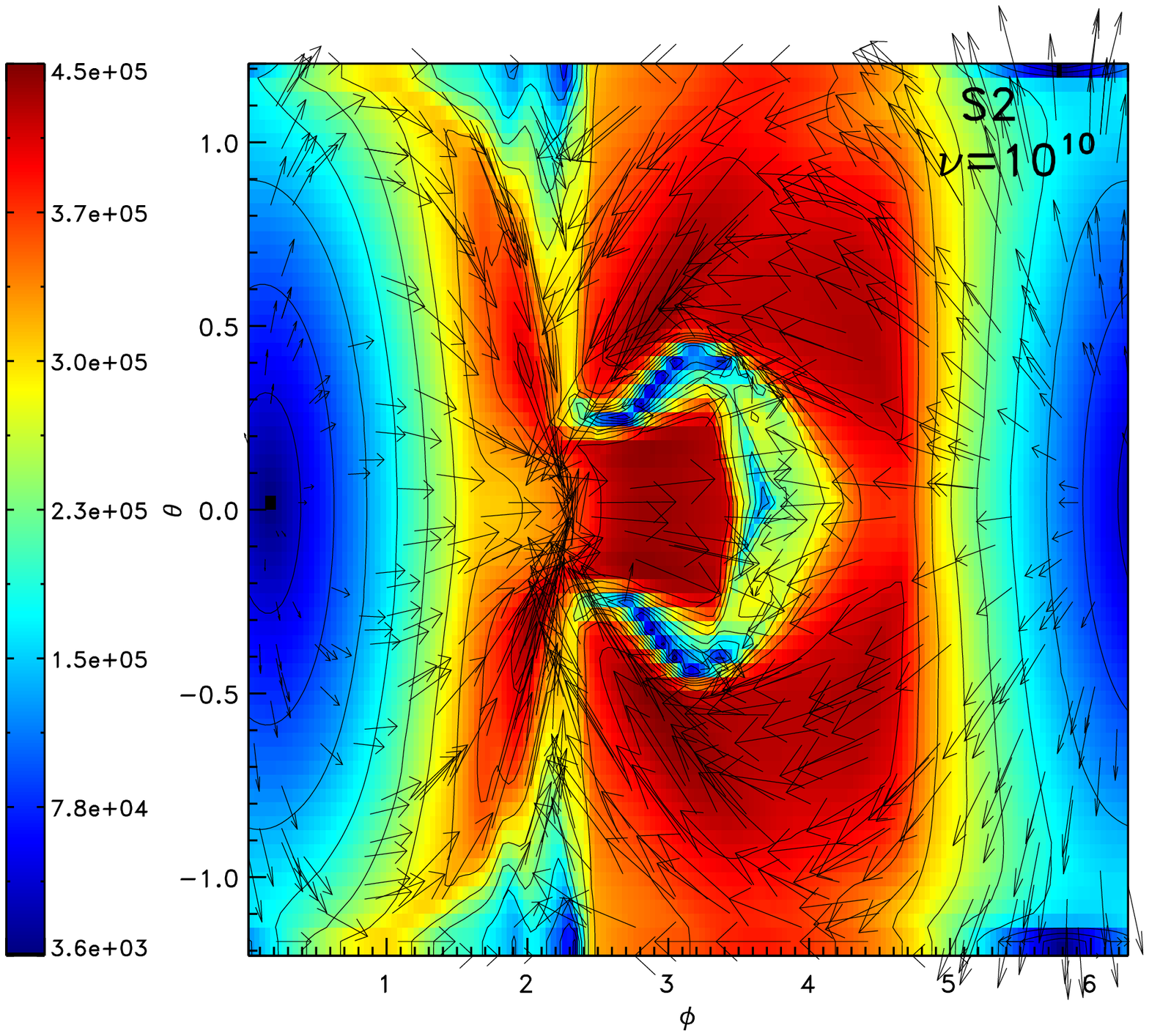}
\includegraphics[height=7.0cm,width=8.0cm]{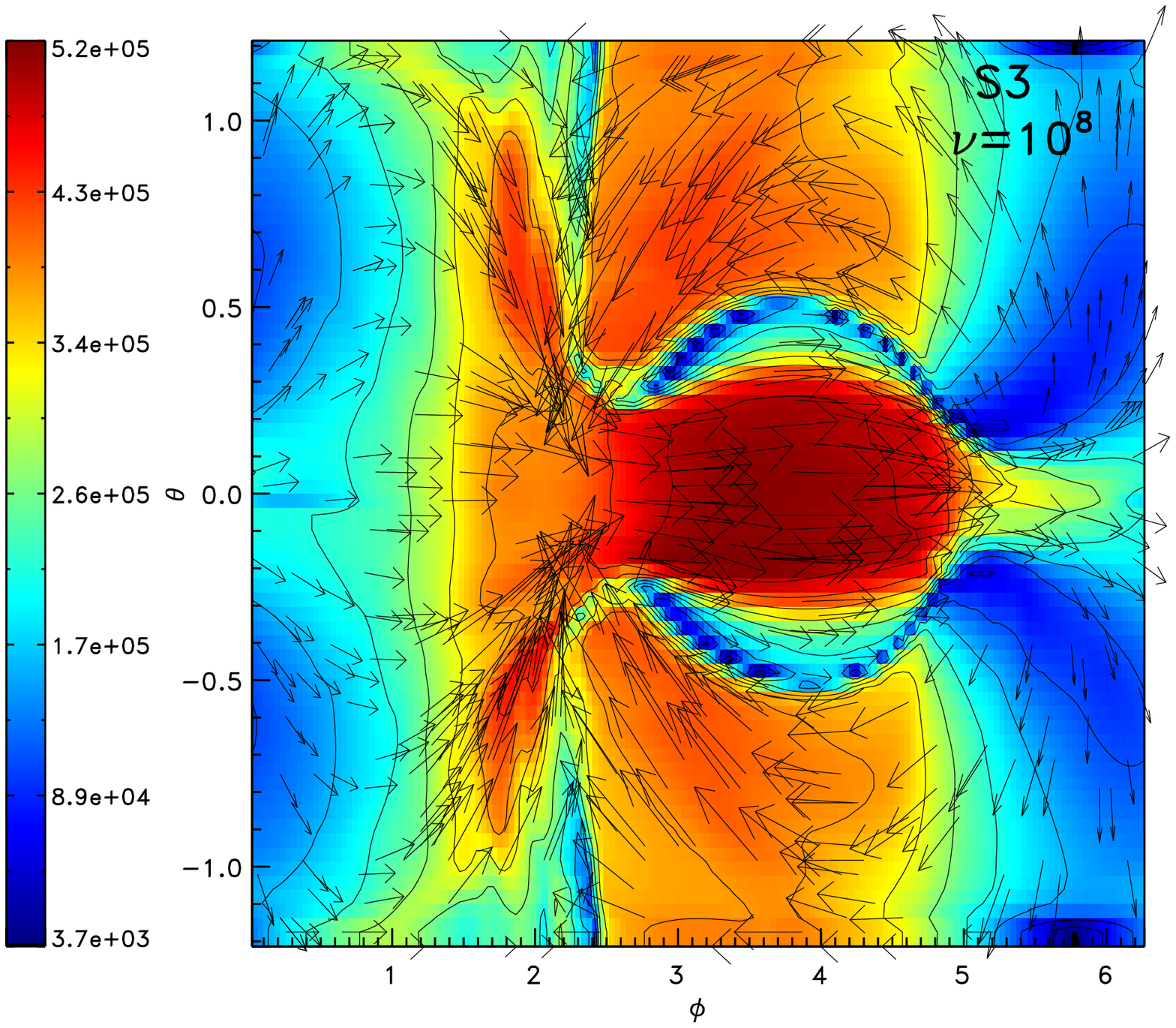}
\hfill
\includegraphics[height=7.0cm,width=8.0cm]{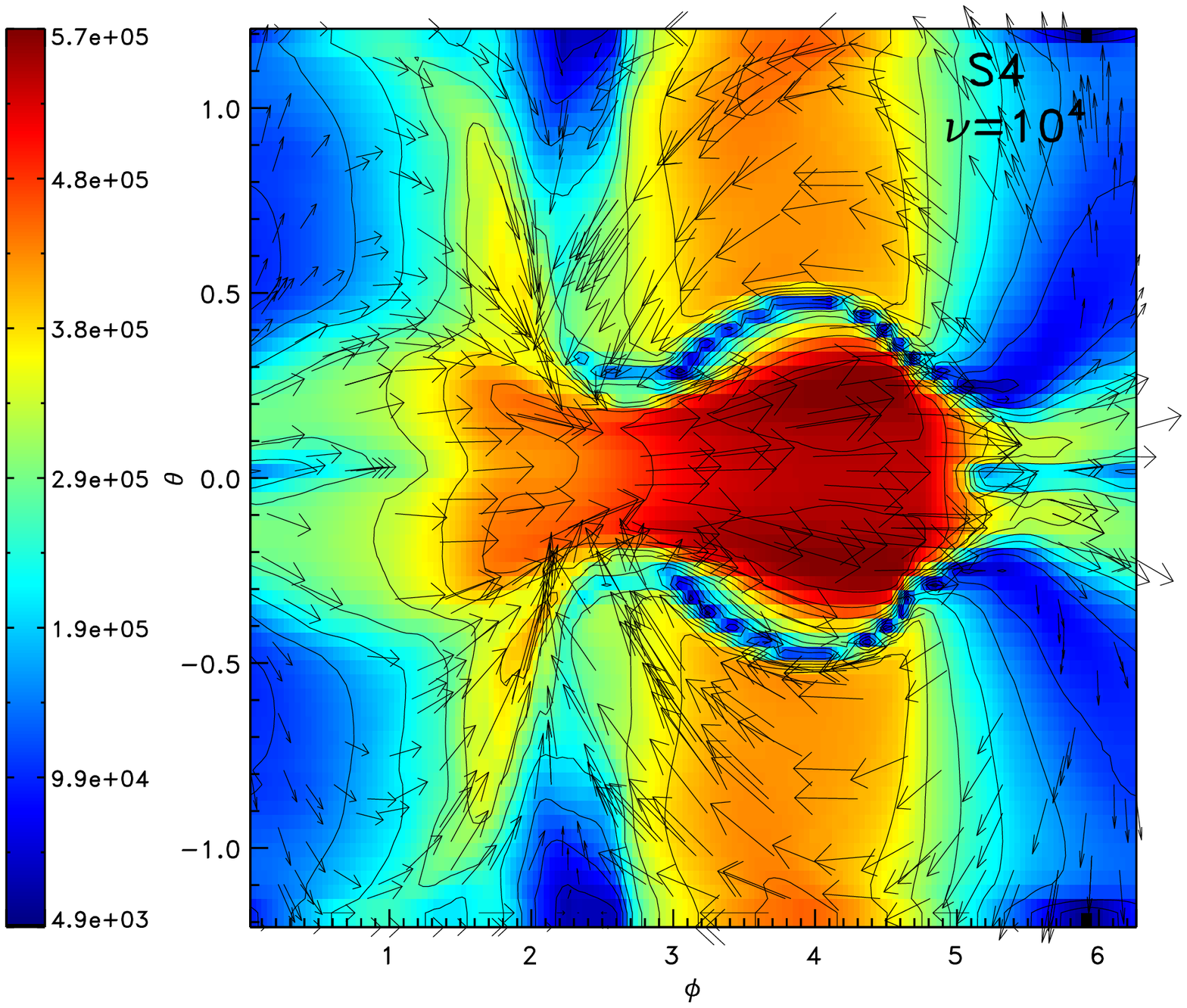}
\caption{The total speed ($v=\sqrt{u_r^2+u_{\phi}^2+u_{\theta}^2}$)
in $\mathrm{cm/s}$ at the photosphere for simulations with varying
viscosities as in Figure (\ref{fig:Tph}). Over-plotted are velocities
across the photosphere, with the arrow length proportional to velocity
magnitude. Very little flow is found for high viscosity, while
supersonic flows are found for the other simulations. The behavior of
the jets also depends on viscosity, exhibiting both circumplanetary
rotation and subduction.}
\label{fig:Vph}
\end{figure}

Figure (\ref{fig:Vph}) illustrates the total speed at the infrared
photosphere. Here again, the simulations exhibit a range of behaviors
for the varying viscosity, with the highest viscosity (S1) keeping
velocities below $1 \mathrm{km/s}$. For the other simulations, as
fluid moves from the sub-stellar point in the eastward direction ({\it
  ie} to the right in the figures) it is funneled toward the equator,
while westward moving material is pushed toward the poles. In an
inertial frame, planet spins from the west toward the east with an
angular frequency $\Omega = 2.1 \times 10^{-5}$ s$^{-1}$.  Eastward
and westward winds with 5 km s$^{-1}$ in the co-rotating frame
introduce a local change in the azimuthal angular frequency $\Delta
\Omega = \pm 7 \times 10^{-6}$ s$^{-1}$ or $\pm 0.3 \Omega$. The
result is the formation of a banded pattern, with jet-like
structures. The exact mechanism for the formation of jets both in
these simulations and in solar system atmospheres remains an
outstanding question. The anti-symmetry appears to be associated with
the rotation of the planet, perhaps related to the asymmetry of the
Coriolis force (For more detailed plots of this phenomena see
\citet{dobbsdixon2008_1}). Non-rotating simulations, presented in
\citet{dobbsdixon2008_1}, show a very symmetric flow pattern despite
large wind speeds. Despite the high velocity flows, the role of
rotation in shaping the flow is not wholly unexpected as the the Rossby
number, given by the ratio of the advective and rotation terms in
Equation (\ref{eq:momentum}), is of order unity throughout most of the
atmosphere in our simulations. Given the highly unstable nature of the
symmetric flow in a non-rotating planet it seems that only a slight
imposed asymmetry is necessary to form the banded structure seen in
Figure (\ref{fig:Vph}).

Westward moving material at high latitudes always reaches a stagnation
point west of the anti-stellar point as it encounters the opposing
flow near $\phi\sim 140^{\circ}$. At this point the fluid subducts
under its counter-part, slowing considerably as it crosses down
isobars to higher pressures and continues around the planet. The
behavior of fluid moving in the eastward direction at the equator
varies with viscosity. For S2, eastward flowing fluid also subducts
under the opposing flow, much in the same manner as the westward
moving fluid. However, for S3 and S4 the fluid is able to form a
circumplanetary equatorial jet at roughly a single pressure
level. Peak velocities, increasing with decreasing viscosity to over
$5 \mathrm{km/s}$ for S4, are found just east of the anti-stellar
point, but quickly decelerate as it passes back up the pressure
gradient to the dayside. For reference the sound-speed at the
photosphere ranges from $1.4 \mathrm{km/s}$ on the nightside to $2.9
\mathrm{km/s}$ on the dayside, implying the flow velocities can reach
Mach $3.3$ in the equatorial jet. Although S3 and S4 form and maintain
a equatorial surface circumplanetary jet, there is still significant
cooling of the flow after it passes the terminator, as can be seen in
Figure (\ref{fig:Tph}).

\begin{figure}
\includegraphics[height=7.0cm,width=8.0cm]{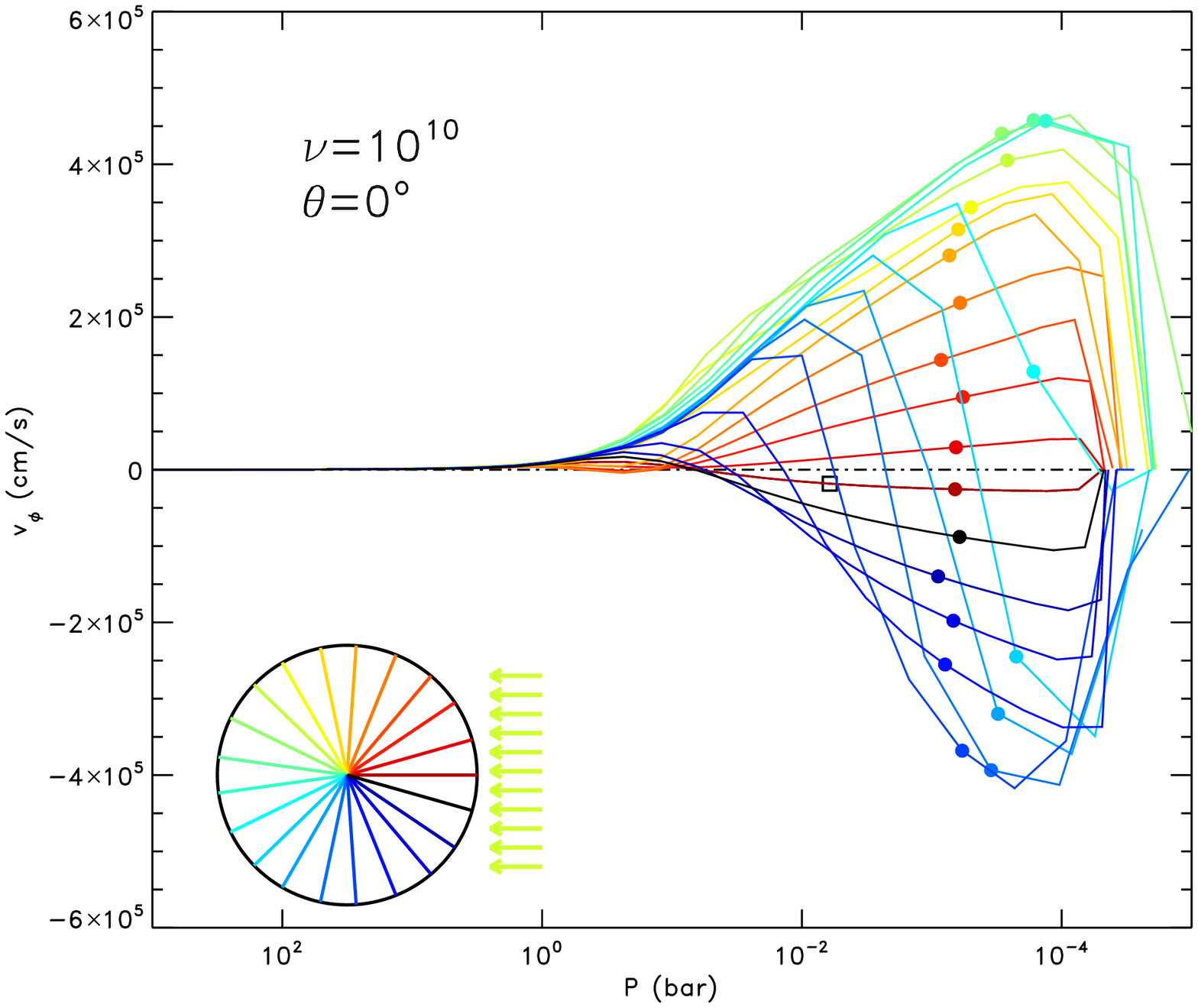}
\hfill
\includegraphics[height=7.0cm,width=8.0cm]{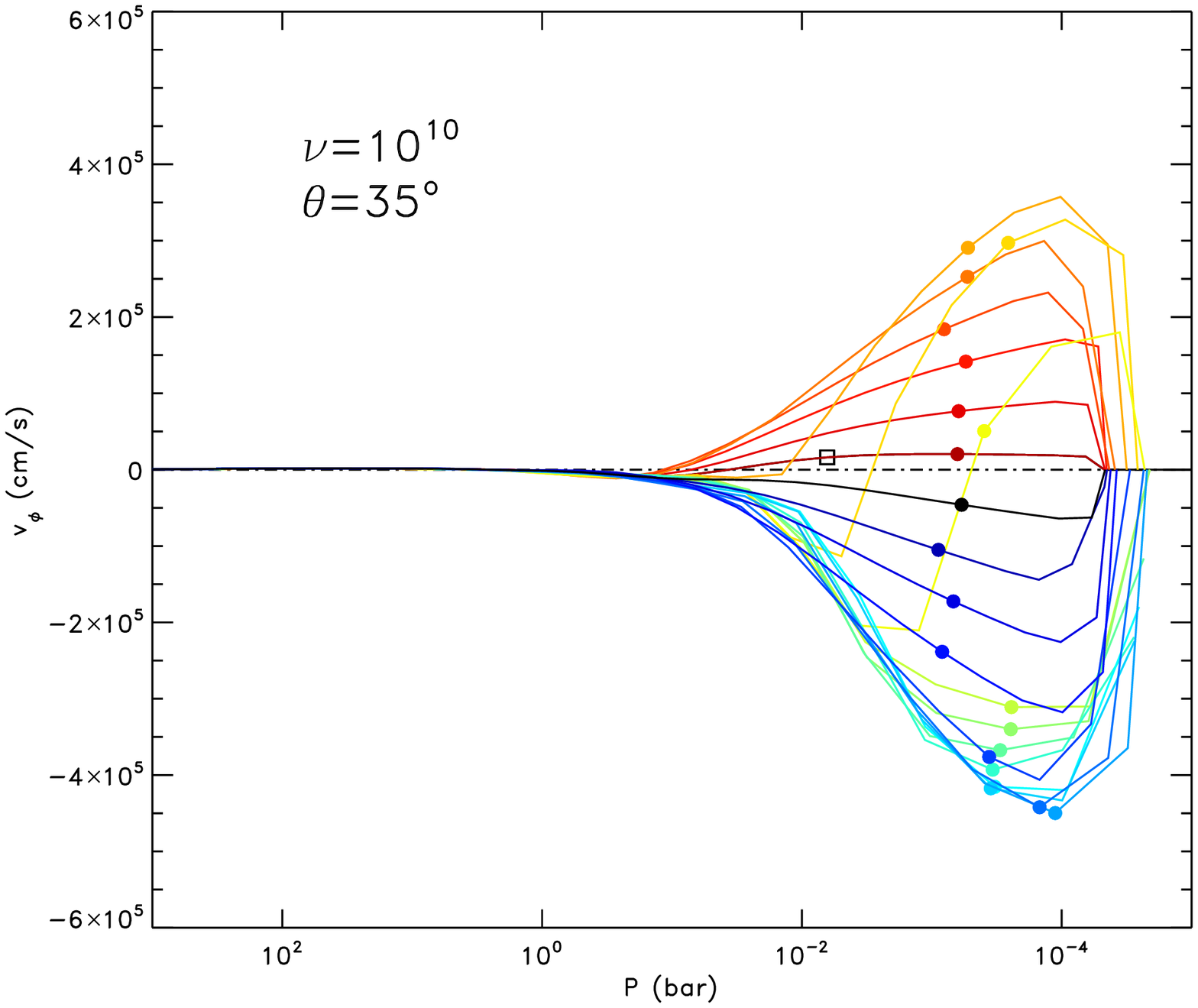}
\includegraphics[height=7.0cm,width=8.0cm]{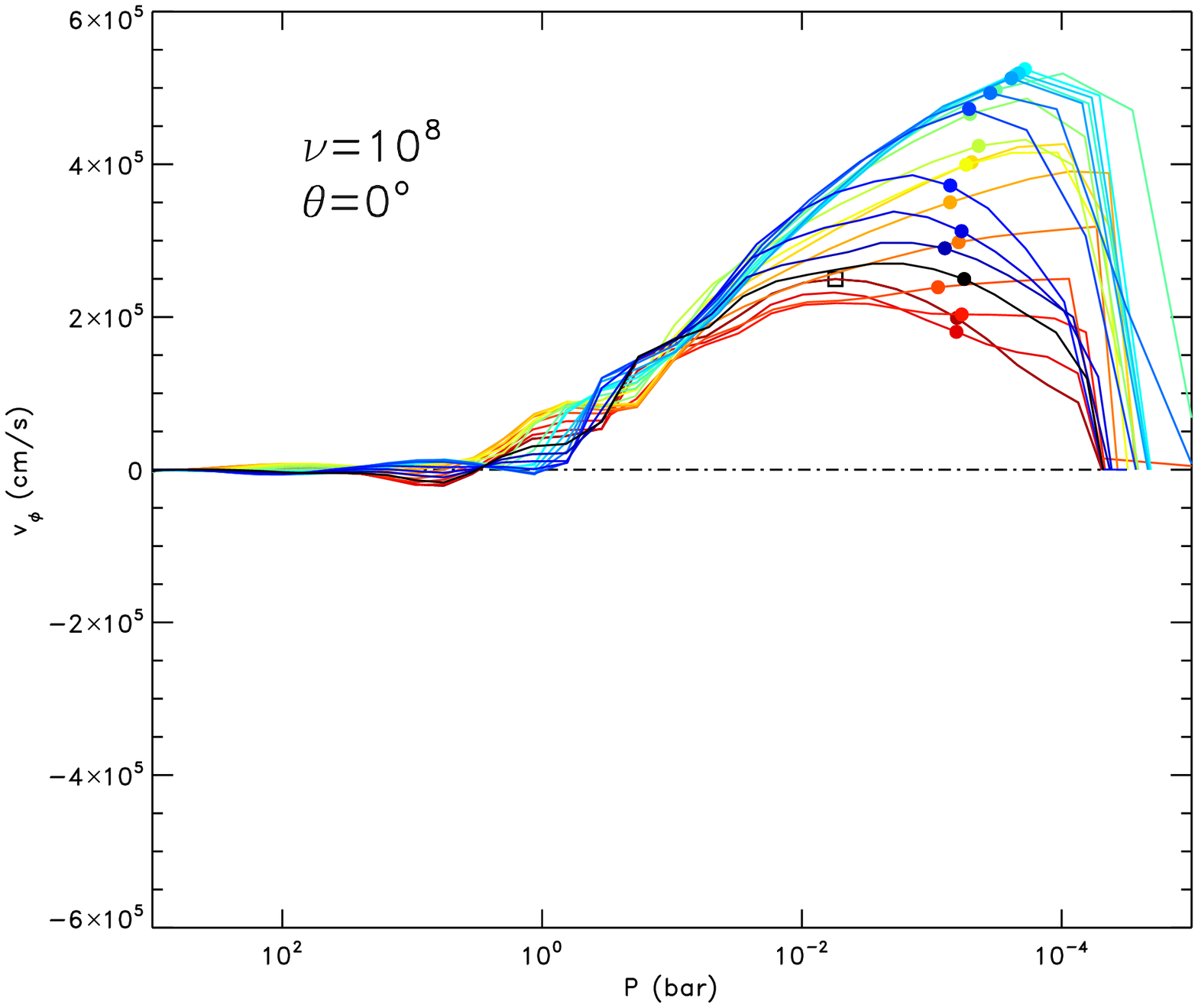}
\hfill
\includegraphics[height=7.0cm,width=8.0cm]{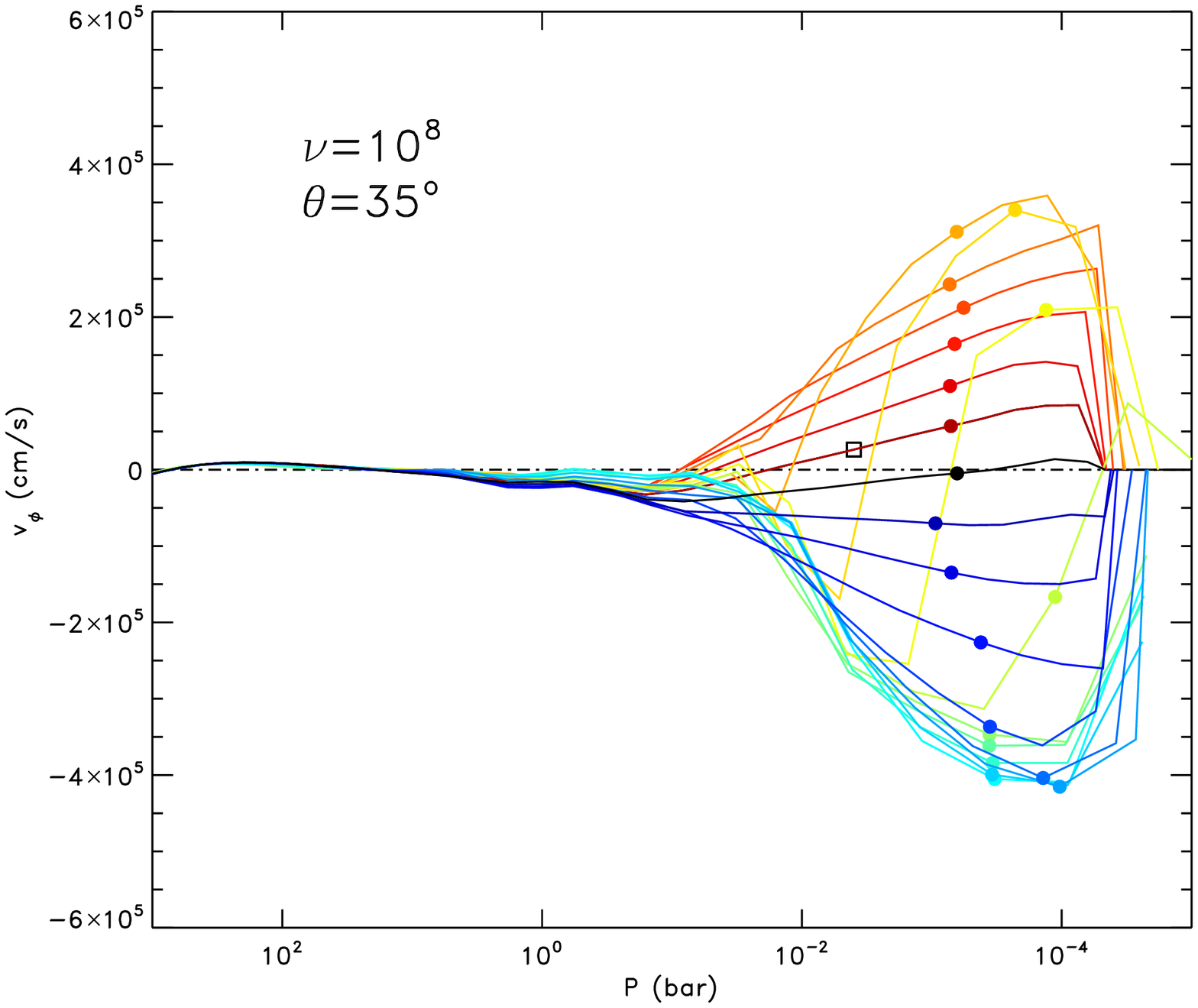}
\caption{The azimuthal velocity ($cm/s$) as a function of pressure at
  the equator for S2 (top row) and S3 (bottom row). The left column is
  a slice at the equator ($\theta=0^{\circ}$) and the right column is
  a slice at mid-latitudes ($\theta=35^{\circ}$). Profiles proceed
  from red at the sub-stellar longitude all the way around the planet
  in $18^{\circ}$ steps in longitude. The key in the upper-left panel
  shows the longitudinal location of a given profile looking down the
  rotation axis, with the arrows indicating the direction of incident
  stellar flux. Solid dots denote the location of the infrared
  photosphere, while the square shows the location of the optical
  photosphere at the substellar point.}
\label{fig:vphip}
\end{figure}

To illustrate the subduction of material at the stagnation points, we
show the azimuthal velocity at the equator ($\theta=0$) and high
latitudes ($\theta=35^{\circ}$) as a function of pressure for S2 and
S3 in Figure (\ref{fig:vphip}). In both simulations, the wind is
launched at low pressures and is accelerated as it passes across the
terminators ($\phi=90^{\circ}$ and $270^{\circ}$). These figures also
clearly illustrate that subduction of fluid occurs for all but the low
viscosity simulations at the equator. For example, in the upper
left-hand plot of Figure (\ref{fig:vphip}) the peak velocity of
eastward moving material (positive $v_{\phi}$) shifts from pressures
near $10^{-4}$ bars down to pressures near $10^{-1}$ bars near the
opposite terminator ($\phi=3\pi /2$). The peak velocities of the
opposing westward fluid remain at low pressures, as the eastward fluid
passes under (see the light blue profile which changes sign at
pressures of approximately $10^{-3}$ bars). The equatorial jet in S3
and S4 extends over a somewhat larger pressure range than the
high-latitude jets or the equatorial jet in S2, and also peaks at
slightly higher velocities. At high latitudes in both S2 and S3 the
opposing streams meet closer to the terminator east of the substellar
point ($\phi=90^{\circ}$). Although some degree of subduction is seen
in the yellow profiles, most of the fluid is pushed toward the
equator, where significant down-welling occurs. The flow toward the
equator can be seen near $\phi=90^{\circ}$ in Figure (\ref{fig:Vph}).

The transition in the overall flow structure (subduction vrs
circumplanetary jet) between S2 and S3 can be related to the critical
value of $\nu_{crit}$ for which viscous effects become comparable with
advection. To quantify this we calculate a modified Reynolds number,
where the length scales used for calculating the viscous and advective
terms are different. We express the crossing timescale as $\tau_{X} =
R_p/u_{avg}$ and the viscous timescale as $\tau_{\nu} = H_p^2/\nu$. We
use the vertical scale-height $H_p$ because we find in our simulations
that the radial transfer of momentum ($\propto \frac{\partial
  u_\phi}{\partial r}$) is overwhelmingly the primary viscous sink for
the jets as they traverse the planet. The ratio of $\tau_{\nu}$ to
$\tau_{x}$ gives our modified Reynolds number, $Re$.  Averaging over
the photosphere we find $H_p\sim3\times 10^{7}cm$, $u_{avg}\sim 1
km/s$, and $R_p\sim 9\times 10^{9}cm$. Setting $Re=1$, we find that
$\nu_{crit}\approx 10^{10} cm^2/s$ is the critical viscosity, above
which viscous forces begin to dominate. For $\nu\ge\nu_{crit}$ viscous
forces try and maintain the symmetry in the flows from both east and
west. The flow patterns for S2 seen in Figure (\ref{fig:vphip})
represents the transition between advection dominated (simulations S3
and S4) and viscosity dominated (simulation S1) flows.

\begin{figure}
\includegraphics[height=7.0cm,width=8.0cm]{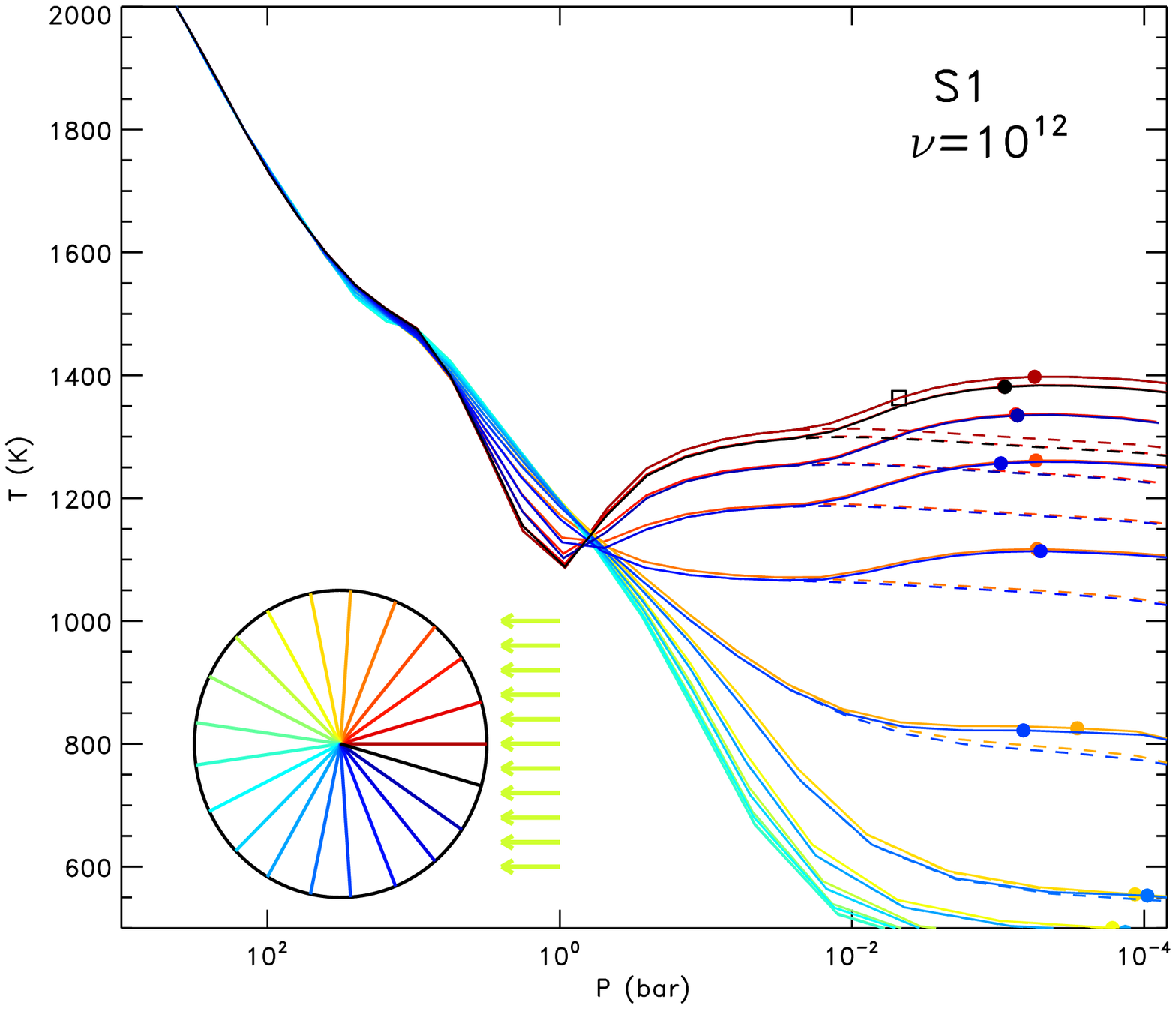}
\hfill
\includegraphics[height=7.0cm,width=8.0cm]{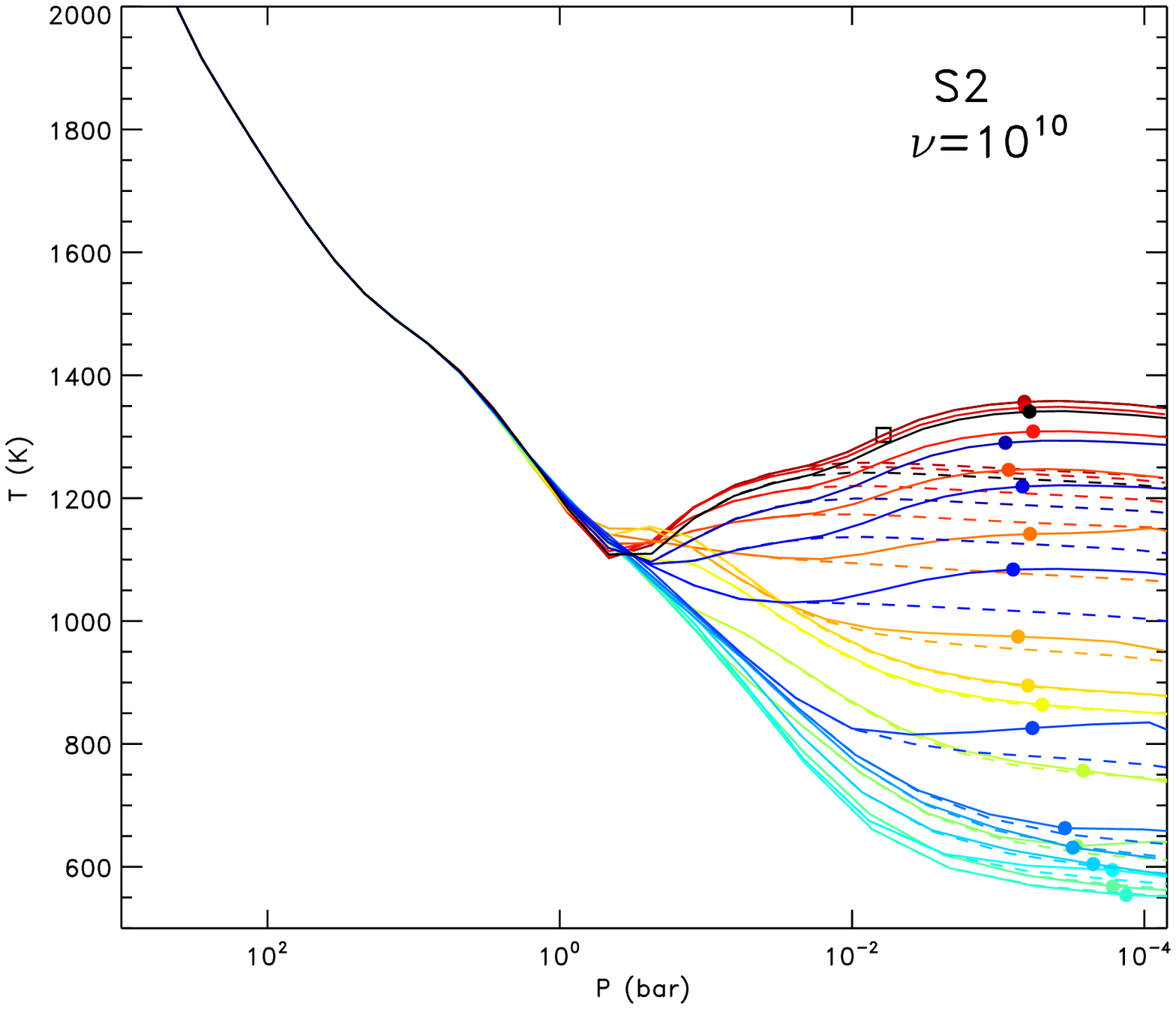}
\includegraphics[height=7.0cm,width=8.0cm]{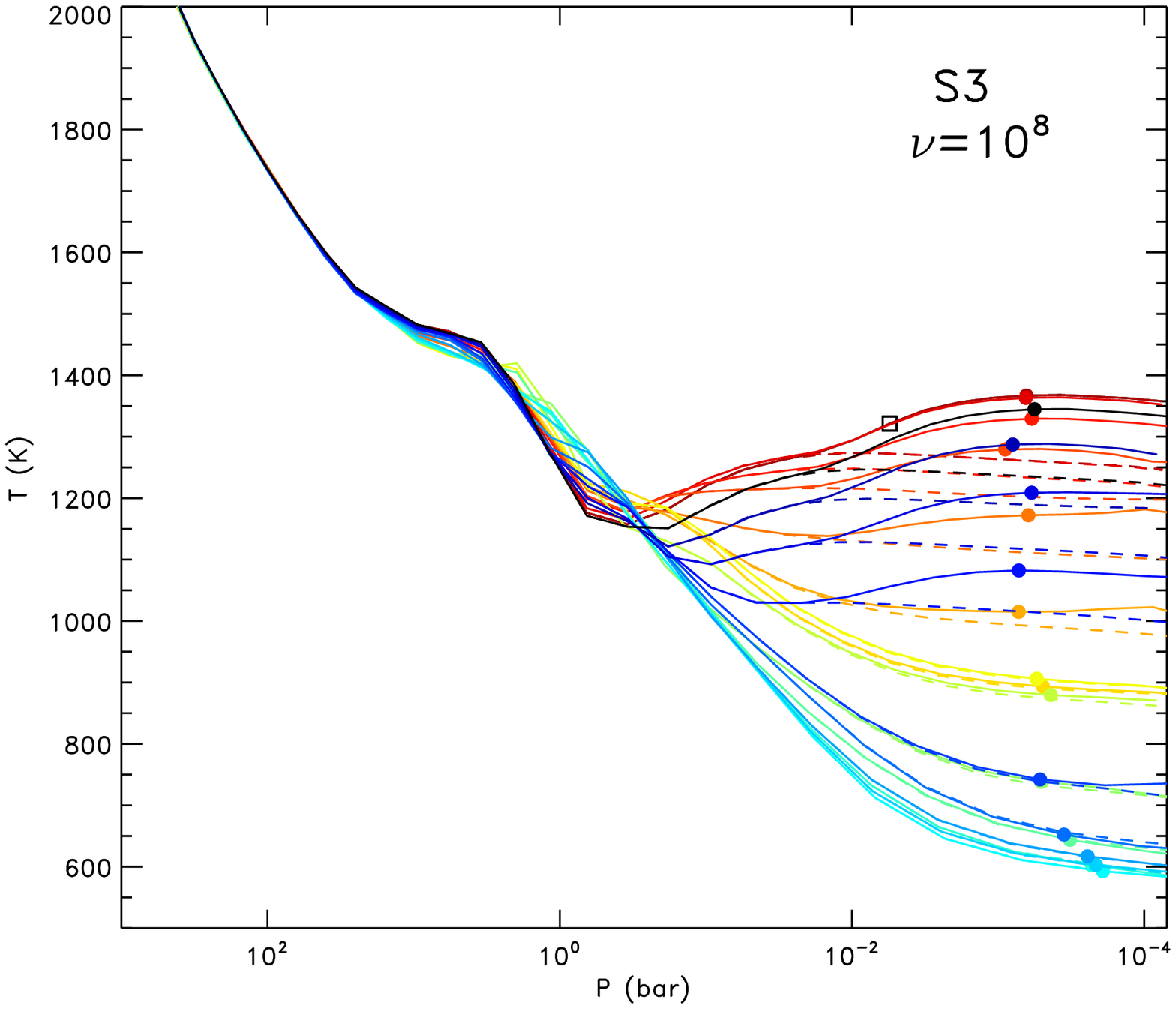}
\hfill
\includegraphics[height=7.0cm,width=8.0cm]{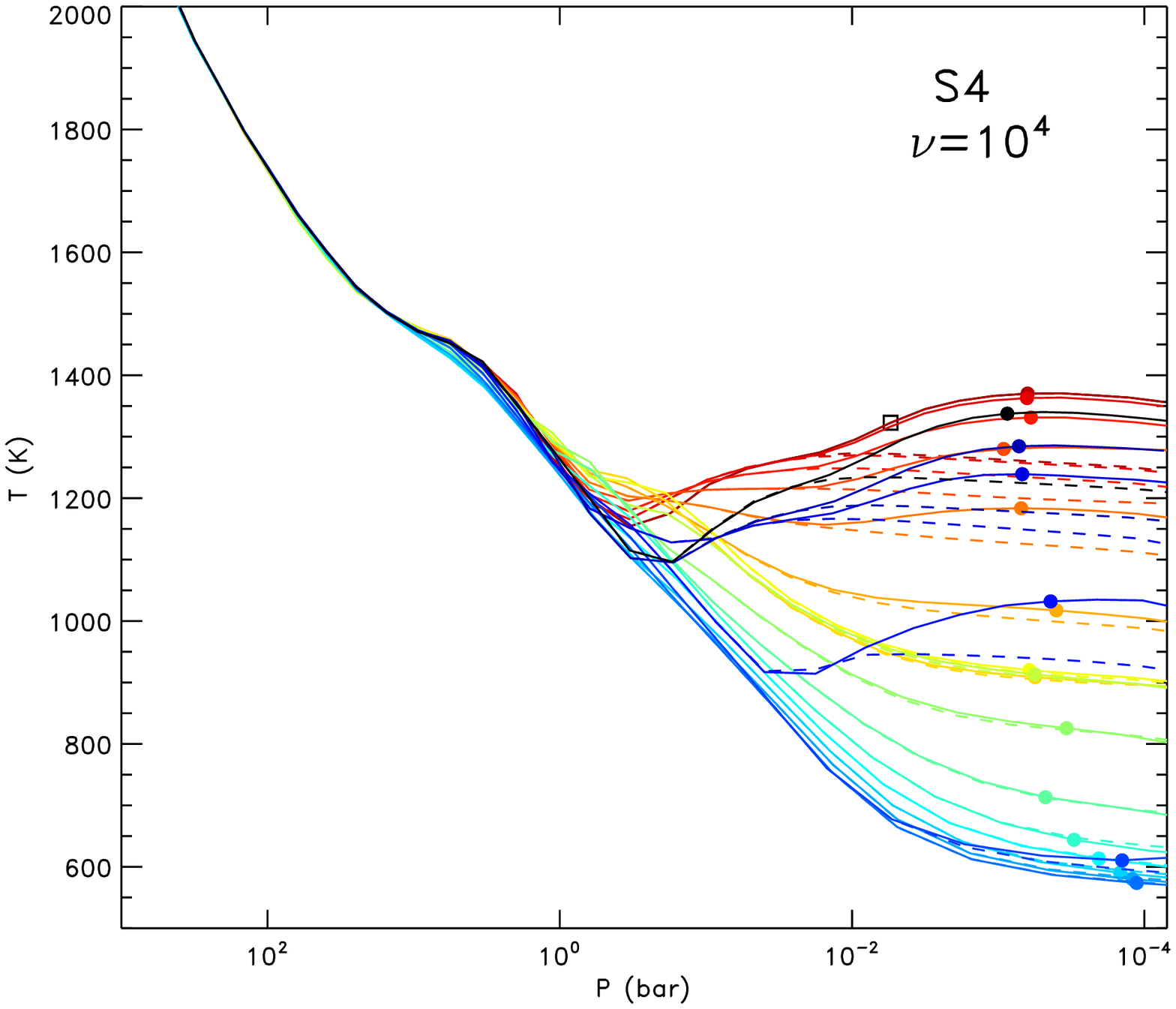}
\caption{Temperature (solid lines) and radiation energy density
$\left(\er/a\right)^{1/4}$ (dashed lines) profiles as a function of
pressure at the equator for S1-S4. Colors and symbols are as in Figure
(\ref{fig:vphip}).}
\label{fig:PT}
\end{figure}

In Figure (\ref{fig:PT}) we show the pressure-temperature profiles at
the equator for all four simulations. The inversion at low pressures
($<10^{-2}$ bars) and interior adiabat are both still clearly evident
as seen in the radiative profile shown in Figure
(\ref{fig:radPT}). Temperatures extend up to $2600$K near the inner
radial boundary, where the planet is spherically symmetric and fully
convecting. Dayside temperature profiles (shown in red) are fairly
consistent across all simulations S2-S4. The upper regions of S1
remain slightly hotter, with temperatures closer to the original
radiative profile of Figure (\ref{fig:radPT}). The larger velocities
calculated in S2-S4 are more effective at advecting away energy,
cooling the dayside. However, the relative similarity between the
dayside photospheric temperature of S2-S4 suggests that the profile of
infrared and optical opacities (Equation (\ref{eq:gamma})) is the
dominant factor in determining dayside temperatures at pressures lower
than $10^{-2}$ bars.

The major effect of the fluid flow on the subsolar profile is at
higher pressures, below both the optical and IR photospheres. The
returning flow significantly cools the atmosphere at pressures around
$1-10^{-2}$ bars creating a secondary, dynamically induced inversion
at depth. The need for two inversions in the pressure/temperature
profile, one caused by the different absorption and emission opacities
and the other caused by the dynamics, was introduced in a
parameterized manner by \citet{burrows2007} in order to simultaneously
match spectral constraints at multiple wavelengths. It arises
self-consistently here. Comparing Figures (\ref{fig:vphip}) and
(\ref{fig:PT}), this cooler region is associated with the flow
\emph{returning} from the nightside after it has circumnavigated the
planet and cooled while on the nightside. The final equilibrium
temperature at this location will be determined by a balance between
the incident energy diffusing downward from the optical photosphere
and the advection bringing in cooler fluid from the nightside. For
this reason, this secondary inversion is slightly warmer and narrower
(in pressure) in S2, than in S3 or S4. As seen above, the higher
viscosity (S2) limits flow velocities, giving the downward diffusing
radiation more time to heat the fluid.

The differences between S2-S4 are most prominent at regions near the
terminator east of the anti-stellar point (see the darker blue profile
in Figure (\ref{fig:PT}). The deep, fast jet of S4 cools the
terminator east of the anti-stellar point ($\phi=3\pi/2$) to a much
greater extent then in any of the other simulations. This cooling
produces a very distinct temperature inversion at this latitude. Such
inversion features should be easily distinguishable in a spectra taken
after primary transit but before secondary eclipse, when this
terminator faces toward earth. In contrast, the terminator west of the
anti-stellar point exhibits no inversion in any of the simulations.

\begin{figure}
\includegraphics[height=6.0cm,width=8.0cm]{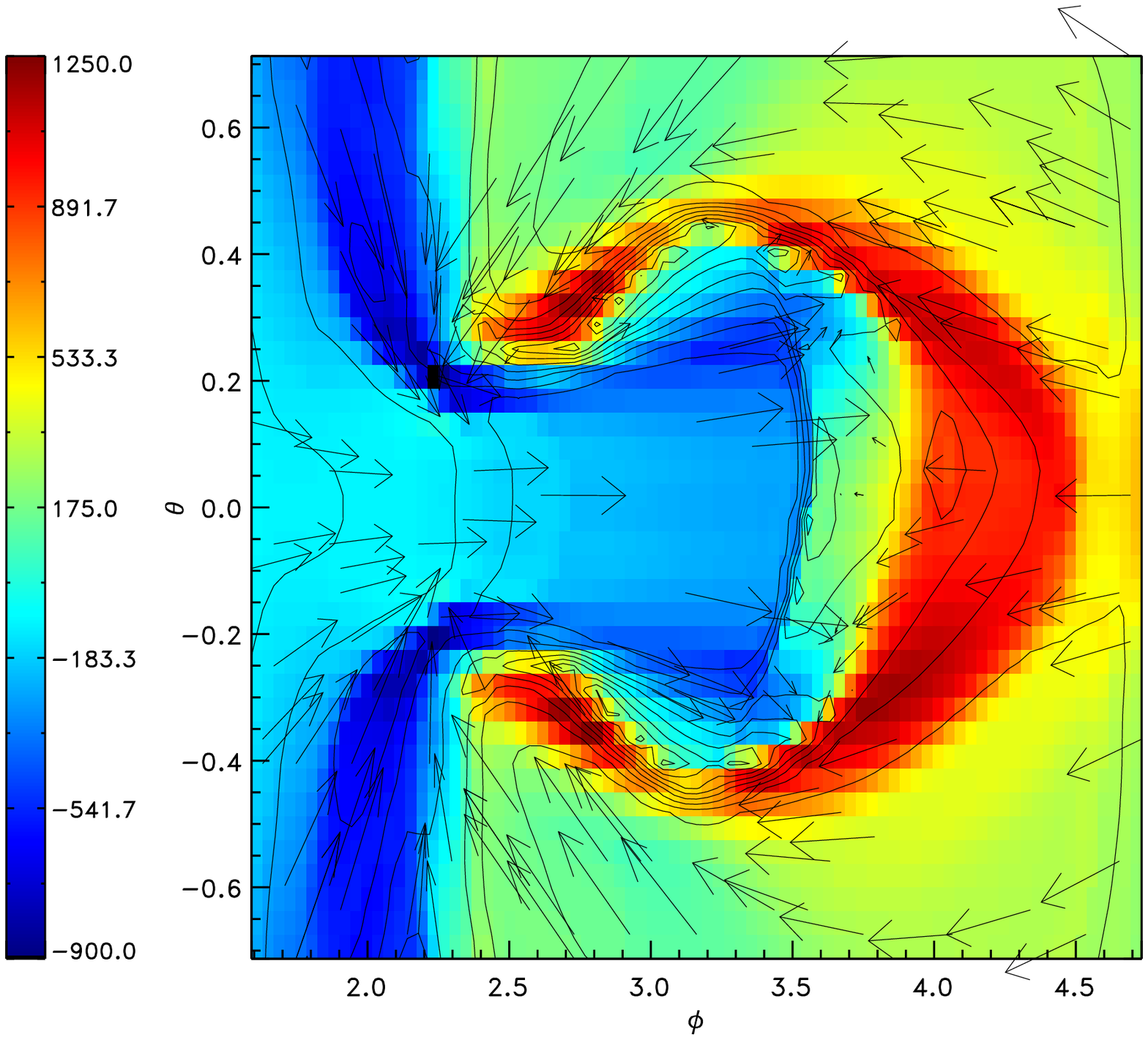}
\includegraphics[height=6.0cm,width=8.0cm]{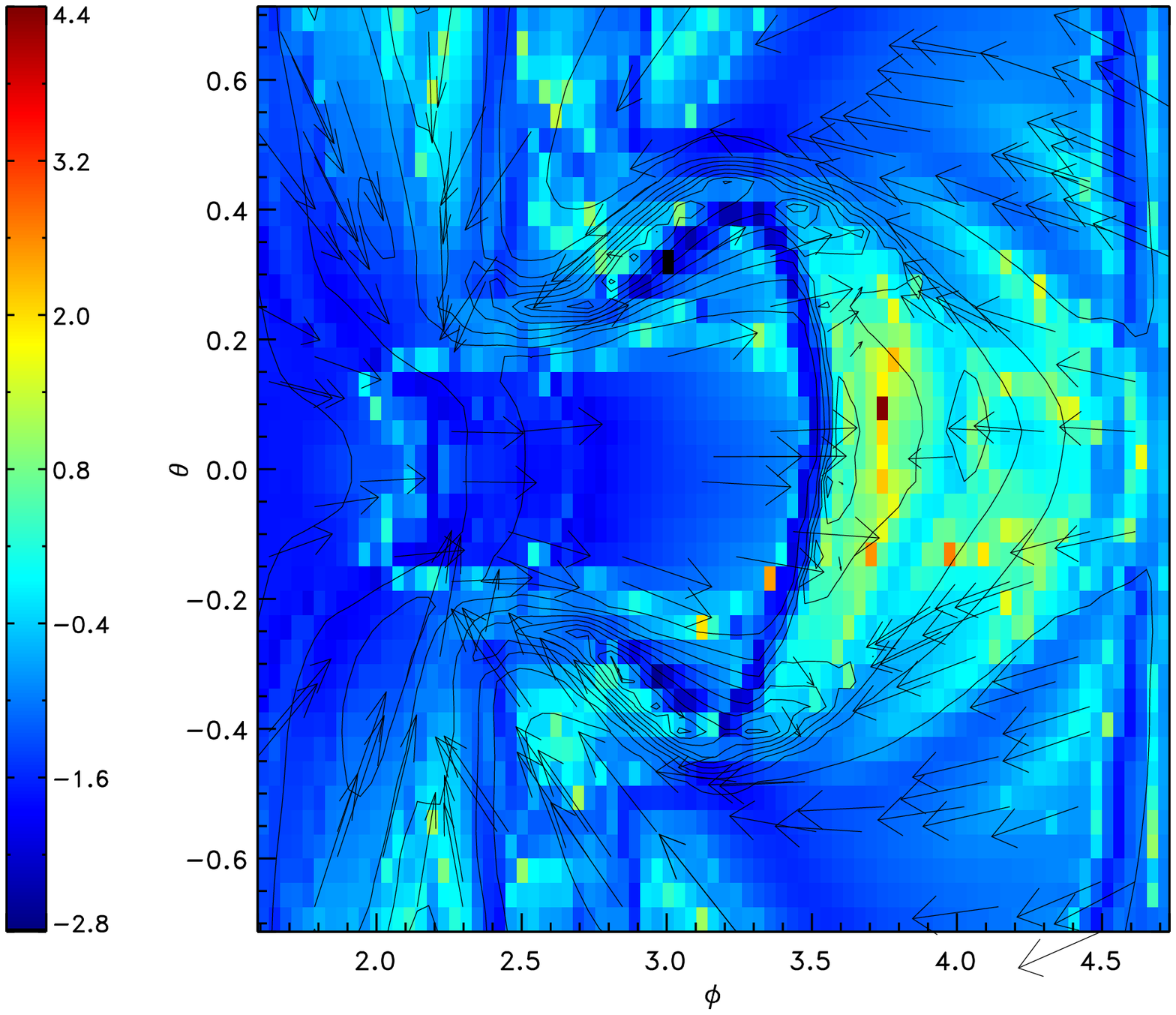}
\caption{The azimuthal component of the viscous force at the
  photosphere (left), acting on $u_{\phi}$. The right panel shows the
  ratio of viscous and compression contributions, plotted as
  $log_{10}\left(\left|D_\nu/P\nabla\cdot{\bf u}\right|\right)$. Both
  graphs are contoured with the total velocity at the photosphere and
  are for S2, with $\nu=10^{10} \mathrm{cm^2/s}$. We have zoomed into
  the region on the nightside centered on the anti-stellar point with
  the equator running horizontally through $\theta=0$, to better
  illustrate this crucial region.}
\label{fig:visc}
\end{figure}

The inclusion of viscosity in the governing equation provides another
mechanism for converting kinetic energy back into thermal energy,
regulating the velocity in the process. The left-hand plot of Figure
(\ref{fig:visc}) illustrates the strength of the viscous term
$\left[\nu\nabla^2{\bf u} +\frac{\nu}{3}\nabla\left(\nabla\cdot{\bf
    u}\right)\right]_{\phi}$ in the longitudinal momentum equation for
S2. This plot concentrates on the convergence region on the nightside,
and clearly illustrates that viscosity primarily acts in this shear
zone. Comparison to Figure (\ref{fig:vphip}) shows that the two
shearing streams have relative velocities of up to $8
\mathrm{km/s}$. The heads of the jets experience the greatest
deceleration, reaching $1250 \mathrm{cm/s^2}$. The importance of
viscous dissipation on the overall energy distribution depends on its
contribution to Equation (\ref{eq:thermalenergy}). The other component
making a major contribution to the thermal energy is the compressional
work, $P\nabla\cdot{\bf u}$. Though for a majority of the atmosphere
$P\nabla\cdot{\bf u}$ dominates, $D_\nu$ becomes more important in the
convergence region for the lower viscosity simulations. In contrast,
for the highest viscosity simulation S1, viscous heating is most
relevant near the terminators where the velocities are highest. The
right-hand plot shows the ratio of the viscous and compression terms
on the nightside near the anti-stellar point. $D_\nu$ exceeds
$P\nabla\cdot{\bf u}$ by several hundred times at the jet heads, which
extend over a wide range in longitude. The details of this convergence
region are paramount in determining the overall dynamical structure
throughout the atmosphere, as it is here where flow either subducts
under its counterpart or instigates circumplanetary jets. Neglecting
the viscous contribution, especially in this region, may lead to
erroneous results. Across all simulations, we find that the average
ratio of $D_\nu$ to $P\nabla\cdot{\bf u}$ near the photosphere roughly
scales with the value of the kinematic viscosity.

The relation between nightside temperature, velocity, and viscosity
is important to understand. Many authors have noted that the
nightside temperature can be understood by equating the crossing
timescale with the radiative cooling timescale; larger velocities
carry more energy to the nightside, decreasing the temperature
differential, and thus the azimuthal acceleration. It is easy to see
an equilibrium will quickly develop. The introduction of viscosity
complicates the issue, controlling the velocity and heating the
gas. Substantial viscous heating requires both shear and large
$\nu$. In S1, the high viscosity prevents flow to the nightside and
thus viscous heating is confined to the dayside. Near the terminators
$D_\nu$ exceeds $P\nabla\cdot{\bf u}$, but the energy is still
dominated by the incident stellar heating, and the viscous
contribution is insignificant. In contrast, there is significant shear
on the nightside of S4, but, due to the low $\nu$, $D_\nu$ is an
order of magnitude smaller. S2 seems to be in a sweet spot. The
viscosity is low enough to allow for shear flows to develop on the
nightside, but high enough to still generate significant heating
there.

\subsection{Time Dependent Behavior: Exo-Weather}
\label{sec:timedep}
\begin{figure}
\includegraphics[height=6.0cm,width=8.0cm]{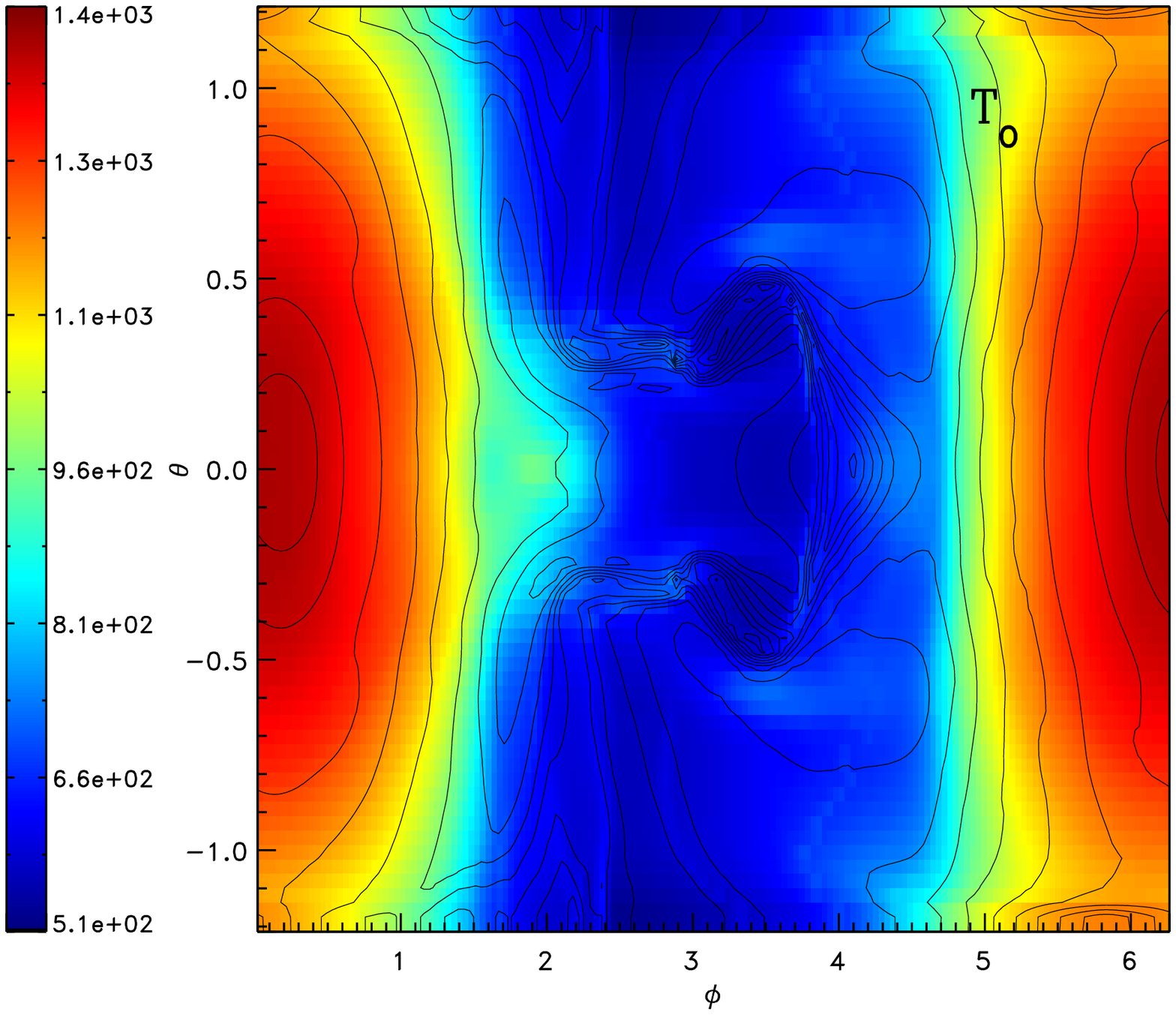}
\hfill
\includegraphics[height=6.0cm,width=8.0cm]{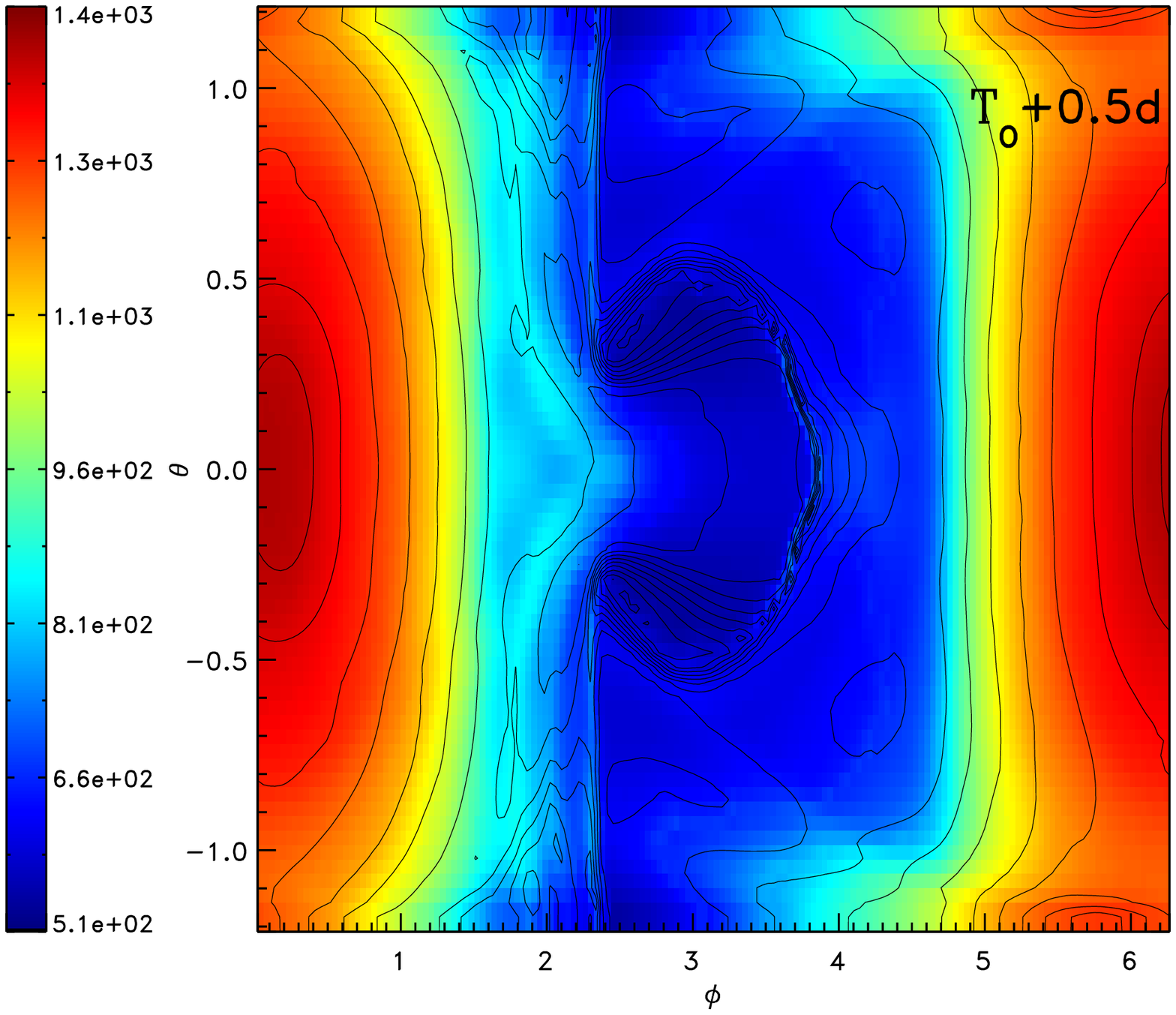}
\includegraphics[height=6.0cm,width=8.0cm]{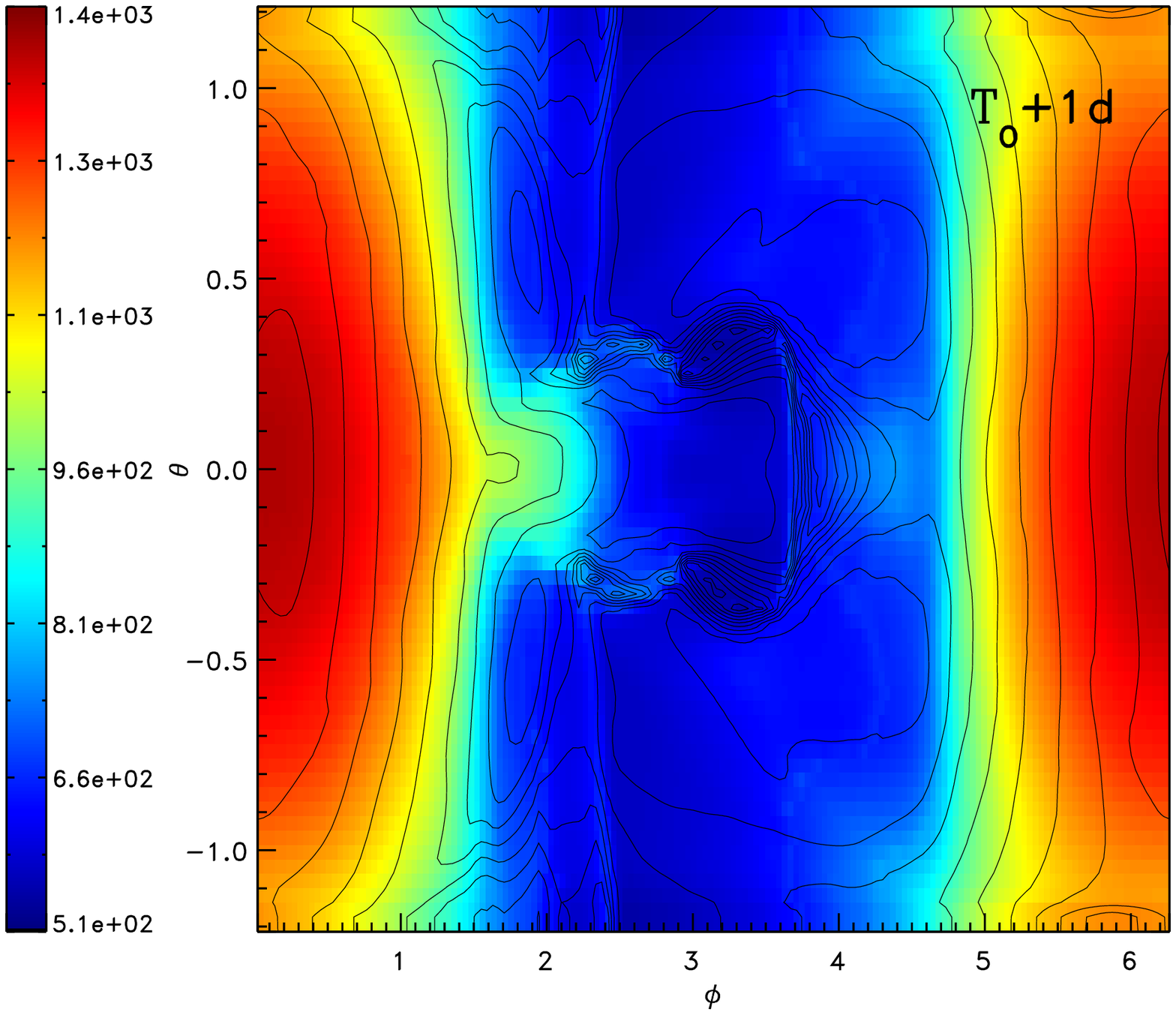}
\hfill
\includegraphics[height=6.0cm,width=8.0cm]{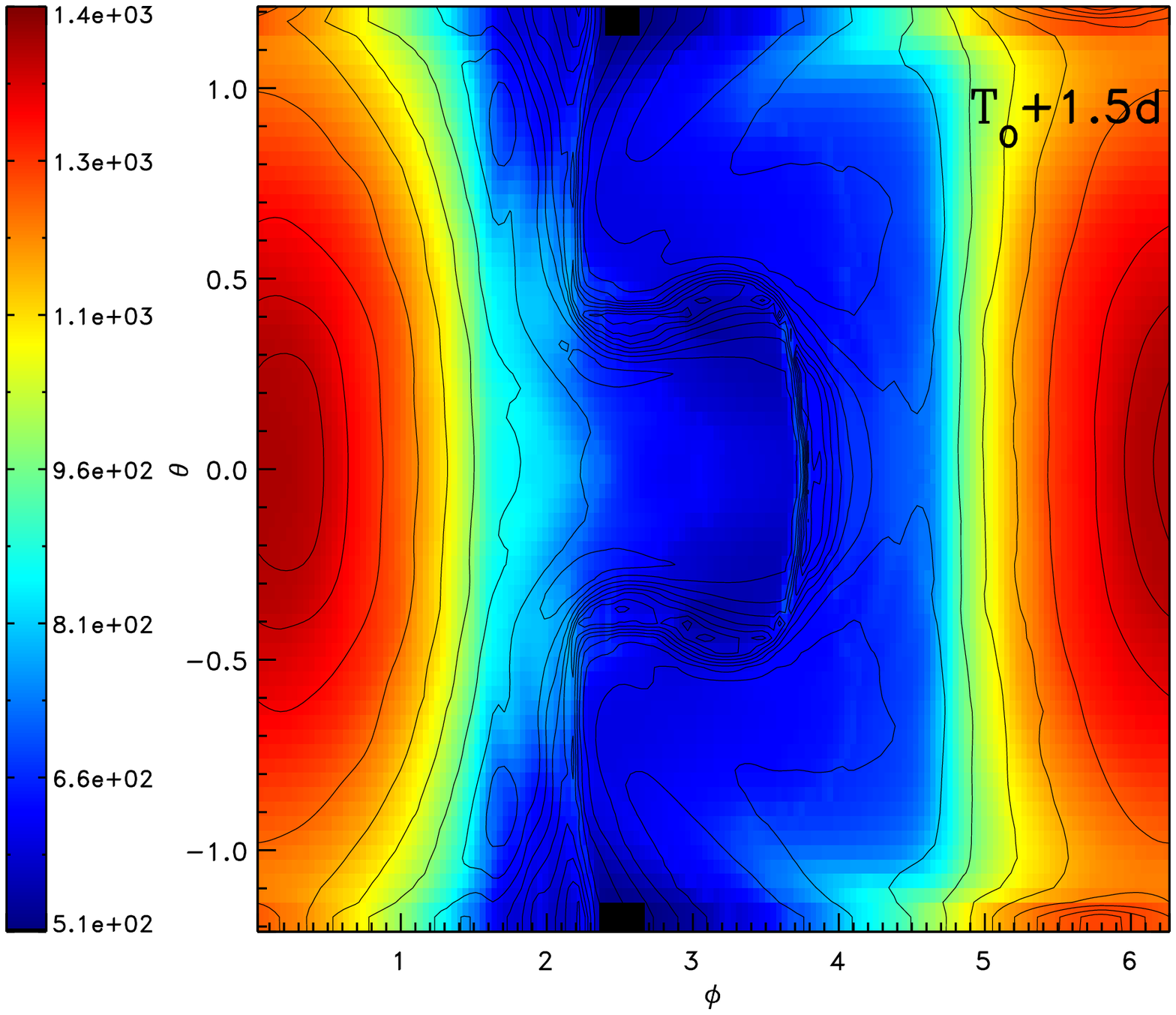}
\caption{The time dependence of S2
  ($\alpha_{eff,ph}=10^{-3}$). Snapshots are shown half a day apart
  and illustrate temperature (color scale) and total velocity
  (contours). Radial and horizontal shearing drives variations in the
  flow, changing the advection of energy across the terminators and
  influencing the location of viscous dissipation.}
\label{fig:timedep}
\end{figure}

Although we have presented average quantities in the Section
(\ref{sec:mean}), the atmosphere can have a dynamic component
super-imposed on the stationary background state. As described above,
the background state consists of a quasi-stationary flow pattern with
flows traveling in both eastward and westward directions. As a result,
sustained supersonic winds traveling in opposite directions converge
on the nightside in the upper atmosphere. Shocks, shearing, and
instabilities result, and the location of the convergence point
oscillates over a range of latitude and longitudes. The shock heating
($D_\nu$) and compressional heating ($P\nabla\cdot{\bf u}$) associated
with this convergence causes the temperature distribution on the
nightside to fluctuate on the timescale of approximately half a
day. Figure (\ref{fig:timedep}) shows the temperature distribution at
the photosphere over a total of $1.5$ days. As can be seen, flow
variations drive temperature changes on the nightside, with the
largest deviations occurring near the terminators.

Changes in the velocity structure are driven primarily by the
interaction between high velocity shearing flows. Shearing occurs both
between streams at constant pressure and also between different
pressure layers. Shearing at constant pressure can be seen on the
nightside in Figure (\ref{fig:Vph}) at $\pm 15^{\circ}$ latitude for
S2-S4 where eastward and westward material pass by each-other and
interact. Shearing between flows at different pressure levels can be
seen in Figure (\ref{fig:vphip}), where the subducting material
travels under counter-flowing material. This can be seen at
$\theta=35^{\circ}$ for S2 and S3, and additionally at the equator for
S2. We find that variations in the temperature at the photosphere are
principally driven by the subduction at the equator, as eastward
moving material sinks below westward moving material at $\phi\sim
3\pi/2$. Significant potential vorticity is generated in this
subduction region, with vortices forming between the two shearing
flows. As the convergence point moves across the surface it forces
these vortices to interact, stirring temperatures further, both at the
photosphere and at depth. Over the course of the entire simulation the
location of the coldest point in the calculated phase-curve is seen to
oscillate more than $20^{\circ}$ in longitude, occasionally shifting
to the west of the anti-stellar point. The largest temperature
variations (up to $15\%$) are seen along the terminators, where
changes to flow patterns have the largest impact on advective
efficiencies.

\section{Enthalpy, Radiation, and Kinetic Energy Fluxes}
\label{sec:fluxes}
One important, overarching questions for groups that model
the atmospheric dynamics of hot-Jupiters is calculating how much
energy is transferred to the nightside, and how changing conditions
change the resulting dynamics and energy distribution. In a recent
article \citet{goodman2008} points out that without viscosity the
atmosphere will adjust to a different solution in order to achieve a
steady state with zero Carnot efficiency. This is fundamentally an
argument regarding the momentum equation, not the energy equation. To
illustrate this, consider the momentum equation (\ref{eq:momentum}) in
steady state, neglecting rotation. Taking the dot product with
velocity and expressing the advective term as $\left({\bf
u}\cdot\nabla\right){\bf u}=1/2\nabla \left|{\bf u}\right|^2+
\left(\nabla\times{\bf u}\right) \times{\bf u}$, we can write
\begin{equation}
{\bf u}\cdot\nabla\left( \frac{1}{2}\left|{\bf u}\right|^2 + w +
\phi_g\right) = {\bf u}\cdot T\nabla S + {\bf u}\cdot\nu\nabla^2{\bf
u} + {\bf u}\cdot\frac{\nu}{3}\nabla\left(\nabla\cdot{\bf u}\right).
\label{eq:bernou}
\end{equation}
The term in parenthesis is the Bernoulli constant ($E_B$) of the flow,
S is the entropy per unit mass, and $w=\epsilon/\rho + P/\rho$ is the
enthalpy per unit mass. Bernoulli's constant is the total of the
kinetic, thermal, and potential energies. Next consider the steady
state form of the two energy Equations (\ref{eq:radenergy}) and
(\ref{eq:thermalenergy}), which together become
\begin{equation}
({\bf u}\cdot\nabla) \epsilon = -P\nabla \cdot {\bf u} -
  \nabla\cdot{\bf F} + S_{\star} + D_\nu.
\end{equation}
$S_{\star} = \rho\kapa F_{\star} e^{-\tau_\star}$ is the direct
stellar energy absorbed by the gas. Utilizing the first-law of
thermodynamics, this can be expressed as
\begin{equation}
{\bf u}\cdot T\nabla S = \rho^{-1}\left[D_\nu - \nabla\cdot{\bf F}
+ S_{\star}\right],
\end{equation}
where the $P\nabla \cdot {\bf u}$ term cancels out, as this does not
contribute to changes in entropy. Plugging this into Equation
(\ref{eq:bernou}) we see that the gradient of $E_B$ along streamlines
is
\begin{equation}
{\bf u}\cdot\nabla E_B = \rho^{-1}\left[D_\nu -
\nabla\cdot{\bf F} + S_{\star}\right] + {\bf u}\cdot\nu\nabla^2{\bf u}
+ {\bf u}\cdot\frac{\nu}{3}\nabla\left(\nabla\cdot{\bf u}\right).
\end{equation}
However, the role of viscosity here is to convert kinetic energy to
thermal energy. Therefore, the viscous contribution from the momentum
equation must be equal and opposite to the viscous contribution from
the energy equation ($\rho^{-1}D_\nu = -{\bf u}\cdot\nu\nabla^2{\bf
u} - {\bf u}\cdot\frac{\nu}{3}\nabla\left(\nabla\cdot{\bf u}\right)$)
leaving us with
\begin{equation}
{\bf u}\cdot\nabla E_B = \rho^{-1}\left[S_{\star} - \nabla\cdot{\bf
F}\right].
\label{eq:entropy}
\end{equation}
Changes in the gradient of $E_B$ along a streamline are solely due to
radiation effects, even when viscosity is included. Since the integral
of ${\bf u}\cdot\nabla E_B$ around a closed streamline must equal zero
in steady-state, so too must the radiation terms. Therefore, the
important difference when including viscosity is the behavior of the
streamlines. Including viscosity will alter the velocity, changing the
streamlines and thus also the overall thermal structure of the
atmosphere. Figures (\ref{fig:Vph}) and (\ref{fig:vphip}) show that
viscosity can play an important role in altering the velocity (and
thus thermal) structures for the planetary atmosphere. Ultimately,
Equation (\ref{eq:entropy}) tells us that the direct effects of
viscosity are important for solving the momentum equation, not the
energy equation, further justifying our choice to solve the full 3D
Navier-Stokes equations.

\begin{figure}
\includegraphics[height=6.0cm,width=8.0cm]{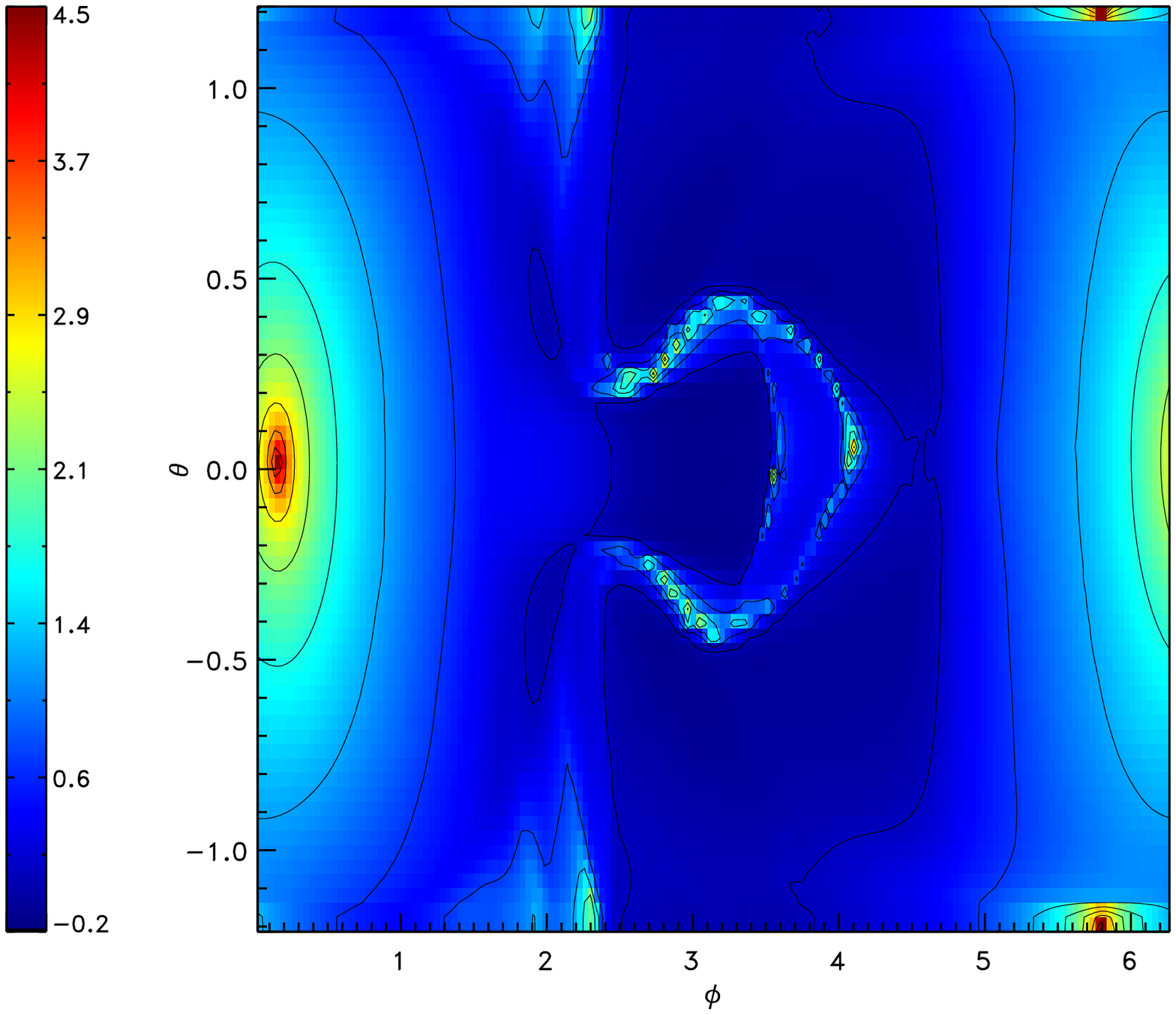}
\includegraphics[height=6.0cm,width=8.0cm]{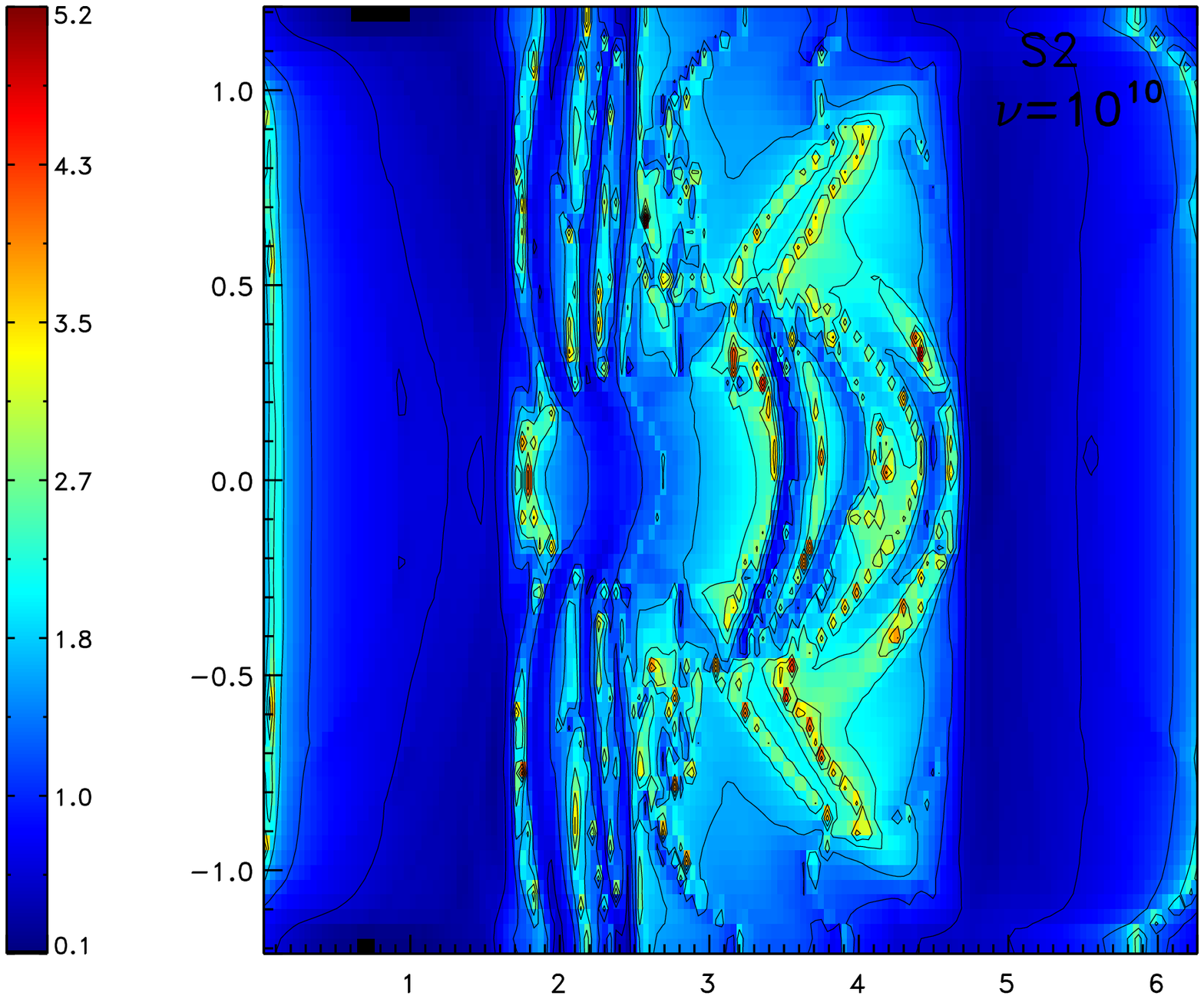}
\caption{The ratio of enthalpy and kinetic fluxes throughout the
  photosphere (left panel). In regions with $F_K>F_E$, it is important
  to include viscous effects. The right panel show the ratio of the
  radial to azimuthal radiative energy flux. Azimuthal radiative
  transport becomes particularly important near the terminators. Both
  plots are shown as the logarithm of the absolute value of the flux
  ratios and are for S2.}
\label{fig:fluxes}
\end{figure}

As pointed out by \citet{goodman2008}, another way to understand the
importance of viscosity is to calculate the relative importance of the
longitudinal energy fluxes carried as kinetic energy vs that carried
as enthalpy. The left-hand panel of Figure (\ref{fig:fluxes}) compares
the azimuthal flux of enthalpy ($\epsilon/\rho + P/\rho$) to the
azimuthal flux of kinetic energy ($\left|{\bf u}\right|^2/2$). Energy
carried in the form of enthalpy dominates that carried in kinetic
energy form throughout most of the interior. However, at higher
altitudes the two contributions become comparable. In particular, in
regions of high velocities on the nightside, where we have already
seen that shocks play a role in determining the dynamical structure,
the flux of kinetic energy exceeds that of enthalpy. In these regions
energy is primarily carried in kinetic form and energy loss through
shocks is important. Our approach of integrating the Navier-Stokes
equations allows us to follow the transfer energy from kinetic to
internal energy across shocks even if the shocks themselves are not
fully resolved. As emphasized by \citet{goodman2008}, this in not the
case for more simplified treatments of the dynamics.

One final advantage of the method we are using here is that radiation
is allowed to flow in all three dimensions. Incident radiation is
modeled here as being purely radial, but re-radiated photons can also
carry energy both to the nightside and pole-ward. Although the radial
component is by far the most important, at high altitudes with
pressures less than $\sim 0.01$ bars, the azimuthal flux of radiation
can become comparable to the radial radiative flux. The ratio of ${\bf
  F}_{r}$ to ${\bf F}_{\phi}$ to is shown in the right-hand plot of
Figure (\ref{fig:fluxes}). Although never exceeding the radial
radiative flux, the azimuthal radiative flux is of the same order
throughout the dayside and extending to the terminators. The effects
of the azimuthal radiative transfer are most prominent near the
terminators, where they play a role in shaping the final
pressure-temperature profiles.

\section{Conclusion}
\label{sec:conc}
We have performed a series of simulations for the atmosphere of the
highly irradiated planet HD209458b. Our approach involves solving the
full 3D Navier Stokes equation coupled to both the thermal and
radiative energy equations. We have separated our frequency averaged
opacities into those relevant for incoming stellar photons and those
for re-radiated photons. This approach allows us to reproduce the
observed temperature inversion in the upper atmosphere and
self-consistently calculate the location that stellar energy is
deposited. This deposited energy drives flow in both the eastward and
westward directions, with eastward flow concentrated near the equator
and westward flow pushed to higher latitudes above and below the
equator. These flows then converge on the nightside and exhibit range
of behavior for various explicit turbulent viscosities.

The viscosity in our simulations spans a wide range, running from
$\nu=10^{12}$ to $10^{4} \mathrm{cm^2/s}$, corresponding to effective
$\alpha_{eff,ph}$-parameters ranging from approximately $10^{-1}$ to
$10^{-9}$. As expected, results vary significantly with viscosity,
especially for higher viscosities. For our highest viscosities,
atmospheric velocities remained subsonic, advection of energy is
suppressed, and the day-night temperature contrast remains high. As
viscosity is decreased flow velocities became super-sonic, reaching
speeds of over $5 \mathrm{km/s}$ in the lowest viscosity
simulation. In addition to the speed, the behavior of the jets
significantly changes with viscosity. For the simulations with low
viscosity ($\alpha_{eff,ph} = 10^{-5}$ and $10^{-9}$) a jet develops
that circumnavigates the planet at a single pressure near the equator.
However for intermediate viscosity ($\alpha_{eff,ph} \approx
10^{-3}$), and at high/low latitudes in all simulations,
circumplanetary flow of a jet can only be achieved accompanied by
subduction. Subduction occurs as the fluid cools and sinks radially,
crossing isobars.

Viscosity also plays a role in determining the variability of the
atmospheric flows and energy distribution. This variability manifests
itself in two ways: temporally varying velocity structures that alter
advection efficiency, and through viscous heating. Seen most
prominently in the simulation with $\nu = 10^{10} cm^2/s$ (S2),
shearing between converging flows on the nightside in both the
horizontal and radial directions drives oscillations in the flow
patterns. These oscillations alter the location of the convergence
point and shape of the streamlines thus changing the rate of energy
advection coming from the dayside, primarily affecting the temperature
near the terminators on the level of $15\%$. Changing streamlines also
shift the region of maximum shear, in turn altering the location of
hot-spots on the nightside induced by viscous heating. The location of
the coldest point in the calculated phase curve can shift by more than
$20^{\circ}$ in longitude, occasionally even appearing west of the
anti-stellar point. Such time-dependent behavior is somewhat
suppressed in our other simulations (S1, S3, and S4). Viscously driven
variations in the flow require that the viscous term ($\nu\nabla^2{\bf
  u}$) in the momentum equation be of order the advective term
($\left({\bf u}\cdot\nabla\right){\bf u}$). This only occurs for
sufficiently high shear (present for low-$\nu$ simulations) and also
large enough kinematic viscosity ($\nu$). The $\nu = 10^{10} cm^2/s$
simulation exists in somewhat of a sweet-spot, allowing high velocity
flows and still maintaining significant viscous heating on the
nightside. Such variation, if observed, may provide a clue to the
nature of the turbulent viscosity in close-in giant planet
atmospheres.

Fluctuations in velocity and temperature are largely confined to the
nightside. Periodic dynamical variations on the dayside are not able
to alter the basic structure at an observable level due to the
dominant incident stellar flux. This is a somewhat different result
then the simulations of \citet{cho2003, cho2008, rauscher2008} and
\citet{menou2008}, where they demonstrate significant variations
throughout the entire phase-curve for a range of selected
parameters. The variations in their models are largely due to cold
circumpolar vortices, though they do observe smaller amplitude
variations near the equator. We do not see such cold circumpolar
vortices, though their development may be impeded due to our
longitudinal boundary conditions. Variations driven by large scale
vortices are considerably different in nature then those presented
here, which arise from the shearing between converging flows and are
confined primarily to the nightside. We find, apart from a modest
downwind advection of the hot-spot, dynamics does very little to alter
the temperature profiles very high in the atmosphere on the
dayside. Simulations with high velocity winds cool the upper
atmosphere slightly, but the final temperatures are fairly similar to
the radiative simulations. The pressure-temperature profile at
pressures less than $10^{-2}$ bars is thus primarily dependent on our
choice of opacity and its behavior with depth as originally shown by
\citet{hubeny2003}. With a lack of better knowledge concerning the
abundance of high opacity sources such as TiO or VO, we have chosen to
augment the opacity with an ad-hoc extra source at high pressures
\citep{burrows2008}. For comparison, the models of HD209458b by
\citet{showman2009} include TiO and VO in the calculation of opacity,
and as a result of this larger $\kapa$ achieve much higher
temperatures at high altitude. Given that this extra opacity is the
dominant factor in determining the temperature, matching future
observations for a wide range of temperatures requires only adjusting
the magnitude of this opacity. Temporal variation in this opacity
would drive variations in dynamics, though \citet{agol2008} studied a
series of secondary eclipse of HD189733b and find that a variation in
$F_{day}/F_{\star}$ of less then $10\%$, consistent with a stable
dayside temperature. Transit spectroscopy appears to be the most
promising observational technique for constraining exo-weather.

The implications of dynamical behavior across the nightside is clear;
it changes the temperature and location of the coldest spot, and it
changes the temperature structure near the terminators. The
concentration of the fluctuations near the equator serves to enhance
the observability of such variations. Average nightside temperatures,
important for constraining interior and evolutionary models, can only
be obtained by averaging a number of such observations. This result
also suggests a possible interpretation for the puzzling results seen
by \citet{knutson2007}, in which they observe a local minimum in the
flux {\em before} the secondary eclipse for HD 189733b. Though we have
no simulated this planet, in light of the dynamical behavior seen in
our simulations the increase in flux before secondary eclipse may be
attributed to transient exo-weather, which shifts the shape of the
planets phase-curve over time.

\acknowledgements We would like to thank Adam Burrows for providing
the opacities, initial 1D models, and useful comments on the
manuscript. We would also like to thank Phil Arras and Eric Agol for
helpful comments on the manuscript. The numerical calculations were
carried utilizing NASA's High End Computing Program computer
systems. This work is partially supported by the Kavli Foundation
which enabled the principle initiation and development of this work at
KIAA-PKU. It is also supported by NASA (NNG06-GH45G, NNX07A-L13G,
NNX07AI88G, NNX08AM4G, HST-AR-11267.01-A), JPL (1270927), and
NSF(AST-0507424). This work was performed in part under contract with
the California Institute of Technology (Caltech) funded by NASA
through the Sagan Fellowship Program and in part by the Canadian
Institute for Theoretical Astrophysics (CITA) National Fellow program.

\bibliographystyle{aa}
\bibliography{ian}

\end{document}